\documentclass[3p]{elsarticle}

\AtBeginDocument{}
\usepackage{cmap}
\usepackage{bm}
\usepackage[T1]{fontenc}
\usepackage{float}
\usepackage{tikz}
\usepackage{amssymb,amsbsy,amsmath,amsfonts,amssymb,amscd}
\usepackage{verbatim}
\usepackage{wrapfig}
\usepackage{algorithmic}
\usepackage{graphicx}
\usepackage{graphics}
\usepackage{stmaryrd}
\usepackage[small]{caption}
\usepackage{subcaption}
\usepackage{hyperref}
\usepackage{color}
\usepackage{amsthm}  
\usepackage{tabularx}
\usepackage{dcolumn}
\restylefloat{table}
\usepackage{booktabs}
\usepackage{xcolor,colortbl}
\usepackage{amssymb}
\usepackage{amsfonts}
\usepackage{graphicx}
\usepackage{parskip}

\newcommand{\norm}[1]{\left|\left|#1\right|\right|}

\hypersetup{
	colorlinks   = true, 
	urlcolor     = blue, 
	linkcolor    = black, 
	citecolor   = black 
}

\graphicspath{{../figures/}}

\begin{document}

\begin{frontmatter}

\title{On the ability of discontinuous Galerkin methods to simulate under-resolved turbulent flows}

\author[add1,add2]{P. Fernandez\corref{1}}
\ead{pablof@mit.edu}
\author[add1,add2]{N.~C. Nguyen}
\ead{cuongng@mit.edu}
\author[add1,add2]{J. Peraire}
\ead{peraire@mit.edu}
\address[add1]{Department of Aeronautics and Astronautics, Massachusetts Institute of Technology, USA.}
\address[add2]{Center for Computational Engineering, Massachusetts Institute of Technology, USA.}
\cortext[1]{Corresponding author}

\begin{abstract}

We investigate the ability of discontinuous Galerkin (DG) methods to 
simulate under-resolved turbulent flows in large-eddy simulation. 
The role of the Riemann solver and the subgrid-scale model in the prediction of a variety of flow regimes, including transition to turbulence, wall-free turbulence and wall-bounded turbulence, are examined. Numerical and theoretical results show the Riemann solver in the DG scheme plays the role of an implicit subgrid-scale model and 
introduces numerical dissipation in under-resolved turbulent regions of the flow. 
This implicit model behaves like a dynamic model and vanishes for flows that do not contain subgrid scales, such as laminar flows, which is a critical feature to accurately predict transition to turbulence. In addition, for the moderate-Reynolds-number turbulence 
problems considered, the implicit model provides a more accurate representation of the actual subgrid scales in the flow than 
state-of-the-art explicit eddy viscosity models, including dynamic Smagorinsky, WALE and Vreman. 
The results in this paper indicate new best practices for subgrid-scale modeling are needed with high-order DG methods.

\end{abstract}


\end{frontmatter}


\section{Introduction}

Over the past few years, discontinuous Galerkin (DG) methods for large-eddy simulation (LES) have emerged as a promising approach to solve complex turbulent flows. DG methods allow for high-order discretizations on complex geometries and unstructured meshes. 
This is critical for the simulation of transitional and turbulent flows in industrial applications, which require that small-scale small-magnitude features are accurately propagated over complex three-dimensional geometries. In addition, DG methods are well-suited to emerging computing architectures, including graphics processing units (GPUs) and other many-core architectures, due to their high flop-to-communication ratio \cite{Abdi:2017,Abdi:2017b}. The use of DG methods for LES is being further encouraged by successful numerical predictions \cite{Beck:14,Fernandez:16b,Fernandez:17a,Fernandez:AIAA:17a,Fernandez:PhD:2018,Frere:15,Gassner:13,Murman:16,Renac:15,Uranga:11,Wiart:15}.

Despite the significant research investment on discontinuous Galerkin methods for large-eddy simulation \cite{Beck:2016,Fernandez:AIAA:17a,Fernandez:nonModal:2018,Manzanero:2018,Moura:17,Winters:2018}, the precise roles of the subgrid-scale (SGS) model (if any) and of the Riemann solver used by the DG scheme in the ability to predict transitional and turbulent flows remains unclear. 
In this paper, we 
investigate the 
ability of discontinuous Galerkin methods 
to simulate a variety of flow regimes, including transition to turbulence, under-resolved\footnote{As is customary, we use the term {\it under-resolved} to refer to simulations in which the exact solution contains scales that are smaller than the grid resolution (the so-called subgrid scales).} wall-free turbulence, and under-resolved wall-bounded turbulence. The Taylor-Green vortex problem \cite{Taylor:37} and the turbulent channel flow \cite{LM:2015} at various Reynolds numbers are considered to that end. 
The remainder of the paper is structured as follows. In Section \ref{s:summaryCases}, we summarize the test problems and studies performed. Numerical results for the Taylor-Green vortex and the turbulent channel flow are presented in Sections \ref{s:TGV} and \ref{s:channel}, respectively. We conclude 
with some remarks 
in Section \ref{s:conclusions}. The details of the DG discretization are presented in \ref{s:HDGdiscretizationApp}.


\section{\label{s:summaryCases}Summary of test cases and studies performed}

We consider the nearly incompressible Taylor-Green vortex \cite{Taylor:37} and turbulent channel flow \cite{LM:2015} problems at various Reynolds numbers. The focus in the Taylor-Green vortex is to investigate the effect of the Riemann solver and the SGS model on the dissipation of kinetic energy. The focus in the turbulent channel flow is to investigate the effect on the turbulent transport. In both test problems, the fluid is assumed to be Newtonian, calorically perfect, in thermodynamic equilibrium, with Fourier's law of heat conduction, and with the Stokes' hypothesis, as discussed in \ref{s:HDGdiscretizationApp}. The dynamic viscosity $\mu$ is constant, the Prandtl number $Pr = 0.71$, and the ratio of specific heats $\gamma = c_p / c_v = 1.4$. The complete description the problems is presented in Sections \ref{s:TGV_caseDescription} and \ref{s:channel_caseDescription}.

We focus on third-order DG methods. 
For this accuracy order, the large scales in the flow are affected less 
by the numerical dissipation of the DG scheme than by the viscous dissipation due to the eddy viscosity in the explicit SGS models considered\footnote{
The numerical dissipation in an $\ell$-th order DG method vanishes at the rate $\mathcal{O}(k^{2 \ell})$ in the small wavenumber limit $k \to 0$ \cite{Ainsworth:2004}. 
The viscous dissipation of a second-order operator, such as the viscous operator of the Navier-Stokes equations, vanishes at the rate $\mathcal{O}(k^{2})$. This is, the decay rate of a signal is proportional to the $2 \ell$-th power of its wavenumber and to the square of its wavenumber, respectively. 
Regarding dissipation at high wavenumbers, the numerical dissipation in high-order DG methods is more localized near the grid Nyquist wavenumber than the viscous dissipation \cite{Fernandez:nonModal:2018}.}. 
A discussion on how our results are expected to extend to higher accuracy orders is presented in Section \ref{s:conclusions}. Due to their lower computational cost for moderately high accuracy orders, we use hybridized DG methods \cite{Fernandez:17a}, a class of discontinuous Galerkin methods that includes the HDG \cite{Nguyen:12,Peraire:10}, the EDG \cite{Peraire:11} and the IEDG \cite{Fernandez:16a} methods. The details of the hybridized DG methods for the compressible Euler and Navier-Stokes equations are presented in \ref{s:HDGdiscretizationApp}. 
The third-order, three-stage $L$-stable diagonally implicit Runge-Kutta DIRK(3,3) scheme \cite{Alexander:77} is used for the temporal discretization, and the time-step size is chosen sufficiently small so that the spatial discretization error dominates the time discretization error \cite{Fernandez:AIAA:17a}.

For each test problem, we perform two studies: One for the Riemann solver and another for the SGS model. For the Riemann solver studies, we consider the following stabilization matrices in the hybridized DG scheme
\begin{subequations}
\label{e:stabMatrices}
\begin{alignat}{1}
\label{e:stabMatrices1} \bm{{{\sigma}}} & \, = \, \frac{1}{2} \big( \bm{A}_n (\widehat{\bm{u}}_h) + | \bm{A}_n (\widehat{\bm{u}}_h) | \big) , \\
\label{e:stabMatrices2}
\bm{{{\sigma}}} & \, = \, | \bm{A}_n (\widehat{\bm{u}}_h) | , \\
\label{e:stabMatrices3}
\bm{{{\sigma}}} & \, = \, \lambda_{max} (\widehat{\bm{u}}_h) \ \bm{I} , 
\end{alignat}
\end{subequations}
where $\bm{A}_n = \partial (\bm{F} \cdot \bm{n}) / \partial \bm{u}$ is the Jacobian matrix of the inviscid flux normal to the element face, $\lambda_{max}$ denotes the maximum-magnitude eigenvalue of $\bm{A}_n$, $| \, \cdot \, |$ is the generalized absolute value operator, $\bm{I}$ is the identity matrix, and $\widehat{\bm{u}}_h$ is the approximation to the trace of the solution on the element faces, as described in \ref{s:HDGdiscretizationApp}. We note that the stabilization matrix implicitly defines the Riemann solver in hybridized DG methods. Additional details on these stabilization matrices and on the relationship between the stabilization matrix and the resulting Riemann solver are presented in \cite{Fernandez:AIAA:17a}. For the purpose of this paper, we note that the two first stabilization matrices lead to Roe-type Riemann solvers, and the third one yields a Lax-Friedrichs-type solver. No explicit SGS models are used for the Riemann solver studies.

For the SGS model studies, we focus on state-of-the-art eddy viscosity models. These models are based on the Boussinesq eddy viscosity assumption and enter the governing equations through an augmented viscous operator. In particular, the static Smagorinsky \cite{Smagorinsky:63}, dynamic Smagorinsky \cite{Lilly:1992}, WALE \cite{Nicoud:1999} and Vreman \cite{Vreman:2004} models are considered, in addition to implicit LES without an explicit model. The explicit models are further equipped with the Yoshizawa model \cite{Yoshizawa:86} for the isotropic part of the SGS stress tensor, the Knight model \cite{Knight:98} for the turbulent diffusion, and an SGS eddy Prandtl number approach with $Pr_e = 0.7$ for the subgrid-scale heat transfer. All other extra terms arising in the Favre-filtered Navier-Stokes equations \cite{Garnier:2009} are not modeled due to their negligible magnitude compared to the previous terms \cite{Martin:2000}. 
The length scale involved in the SGS models is set to $\Delta = V^{1/3} / k$, where $V$ is the volume of the element and $k$ the polynomial order of the DG approximation. For the dynamic Smagorinsky model, projection onto the space of polynomials of degree $k' = \lfloor k / 2 \rfloor$ is used for the coarse-graining, where $\lfloor k / 2 \rfloor$ is the greatest integer less than or equal to $k / 2$, i.e.\ $k' = 1$ in our case. 
The stabilization matrix \eqref{e:stabMatrices1} is used for the SGS studies.

Compressibility effects will be neglected for the two following purposes: (i) Incompressible DNS results will be used as reference data, and (ii) spatial filtering and Favre filtering will be assumed to be equivalent in the presentation of the numerical results. We finally note that, for some of the Reynolds numbers considered, the mesh resolution is intentionally insufficient to match the DNS data, thus allowing for a more meaningful analysis of the role of the implicit and explicit models in under-resolved simulations.

\section{\label{s:TGV}Taylor-Green vortex}

\subsection{\label{s:TGV_caseDescription}Case description}

The Taylor-Green vortex (TGV) \cite{Taylor:37} is a canonical problem in fluid mechanics commonly used to study vortex dynamics, turbulence transition, and turbulence decay. It contains several flow regimes in a single construct and is therefore an excellent test case for our purpose. In particular, the TGV problem describes the evolution of the flow in a cubic domain $\Omega = [-L \pi, L \pi)^3$ with triple periodic boundaries, starting from the smooth initial condition
\begin{equation}
\label{initialCondTGV}
\begin{split}
\rho & = \rho_0 , \\
u & = V_0 \sin \Big( \frac{x}{L} \Big) \cos \Big( \frac{y}{L} \Big) \cos \Big( \frac{z}{L} \Big) , \\
v & = - V_0 \cos \Big( \frac{x}{L} \Big) \sin \Big( \frac{y}{L} \Big) \cos \Big( \frac{z}{L} \Big) , \\
w & = 0 , \\
p & = P_0 + \frac{\rho_0 \, V_0^2}{16} \, \bigg( \cos \Big( \frac{2x}{L} \Big) + \cos \Big( \frac{2y}{L} \Big) \bigg) \, \bigg( \cos \Big( \frac{2z}{L} \Big) + 2 \bigg) , 
\end{split}
\end{equation}
where $\rho$, $p$ and $\bm{v} = (u, v, w)$ denote density, pressure and the velocity vector, respectively, and $\rho_0 , \, P_0 , \, V_0 > 0$ are some reference density, pressure and velocity magnitude. The large-scale eddy in the initial condition leads to smaller and smaller structures through vortex stretching. Near $t \approx 7 \, L / V_0$, the vortical structures undergo structural changes, and around $t \approx 8 - 9 \, L / V_0$ they break down and the flow transitions to turbulence\footnote{
Note that no temporal chaos (chaotic attractor) exists in the viscous Taylor-Green vortex since the flow eventually comes to rest due to viscous dissipation. We use the term {\it turbulence} here to refer to the phase of spatial chaos (spatial decoherence) that takes place after $t \approx 8 - 9 \, L / V_0$ for Reynolds numbers above about $1000$ \cite{Brachet:1983}.
}, where the exact times depend on the Reynolds number $Re = \rho_0 V_0 L / \mu$. 
The Reynolds numbers $1600$ and $\infty$ are considered in this paper. For the $Re = 1600$ case, the smallest turbulent structures and the maximum dissipation rate of kinetic energy occur at $t \approx 9 \, L / V_0$. After this time, the turbulent motion dissipates all the kinetic energy and the flow eventually comes to rest through a decay phase similar to that in decaying homogeneous isotropic turbulence, yet not isotropic here. At this Reynolds number, the subgrid scales are moderate compared to the resolved scales. In the inviscid TGV, however, the smallest turbulent scales become arbitrarily small, and the range of subgrid scales arbitrarily large, as time evolves. 

The reference Mach number is set to $M_0 = V_0 / c_0 = 0.1$, 
where $c_0$ denotes the speed of sound at temperature $T_0 = P_0 / (\gamma - 1) \, c_v \, \rho_0$. 
This completes the non-dimensional description of the problem.


\subsection{Details of the numerical discretization}

The computational domain is partitioned into a uniform $64 \times 64 \times 64$ Cartesian grid and the third-order Embedded DG (EDG) scheme \cite{Peraire:11} is used for the spatial discretization. The solution is computed from $t_0 = 0$ to $t_f = 15 \, L / V_0$. 
Three different phases exist 
in the simulation. Before $t \approx 4 \, L / V_0$, the flow is laminar and with no subgrid scales. This is followed by an under-resolved laminar phase (i.e.\ with subgrid scales) that lasts until $t \approx 7-9 \, L / V_0$. From then on, the flow is turbulent and under-resolved.

\subsection{Numerical results}


\subsubsection{\label{s:RiemannTGV}Riemann solver study}

\begin{figure}[t!]
\centering
\includegraphics[width=0.49\textwidth]{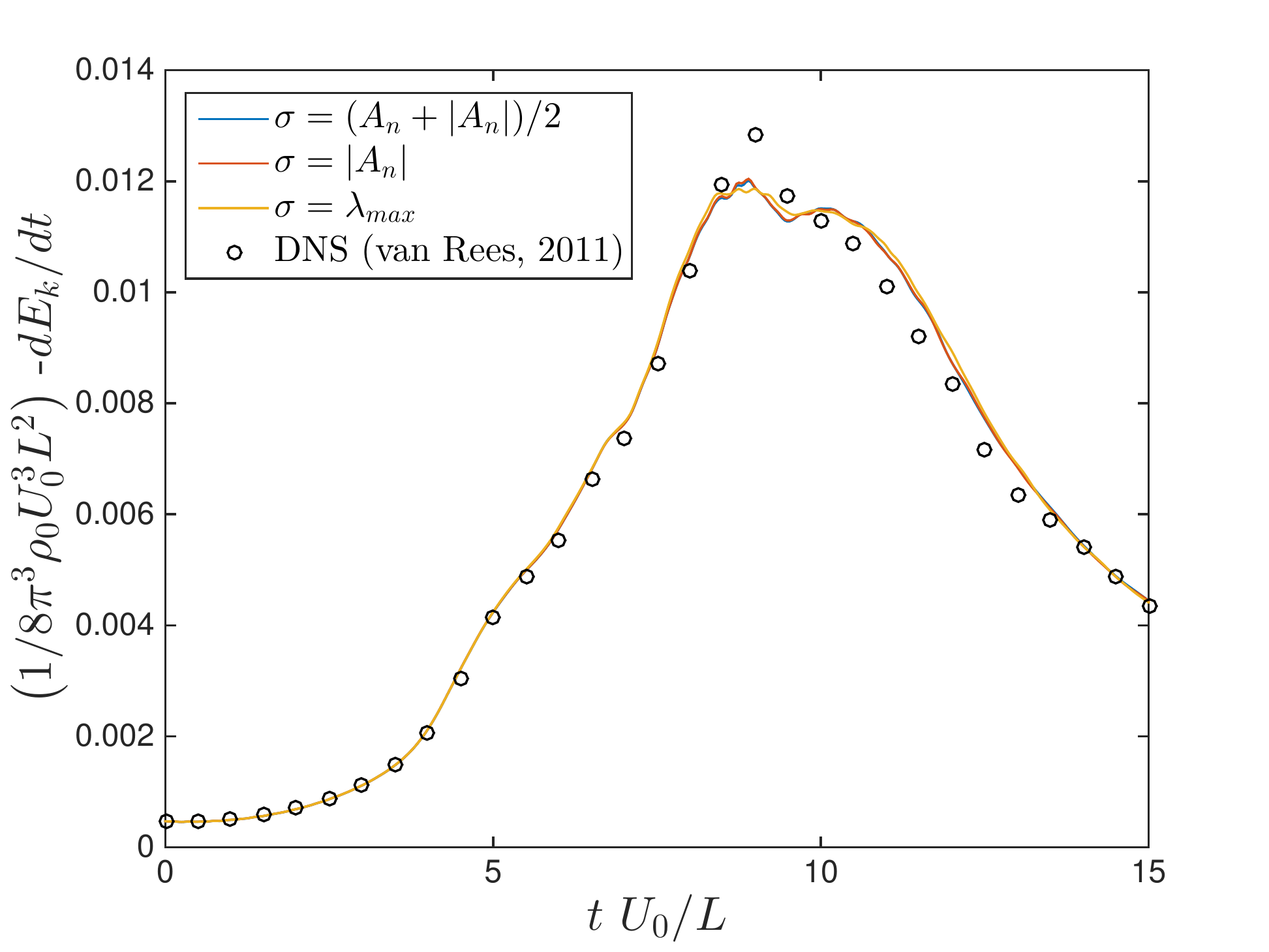}
\includegraphics[width=0.49\textwidth]{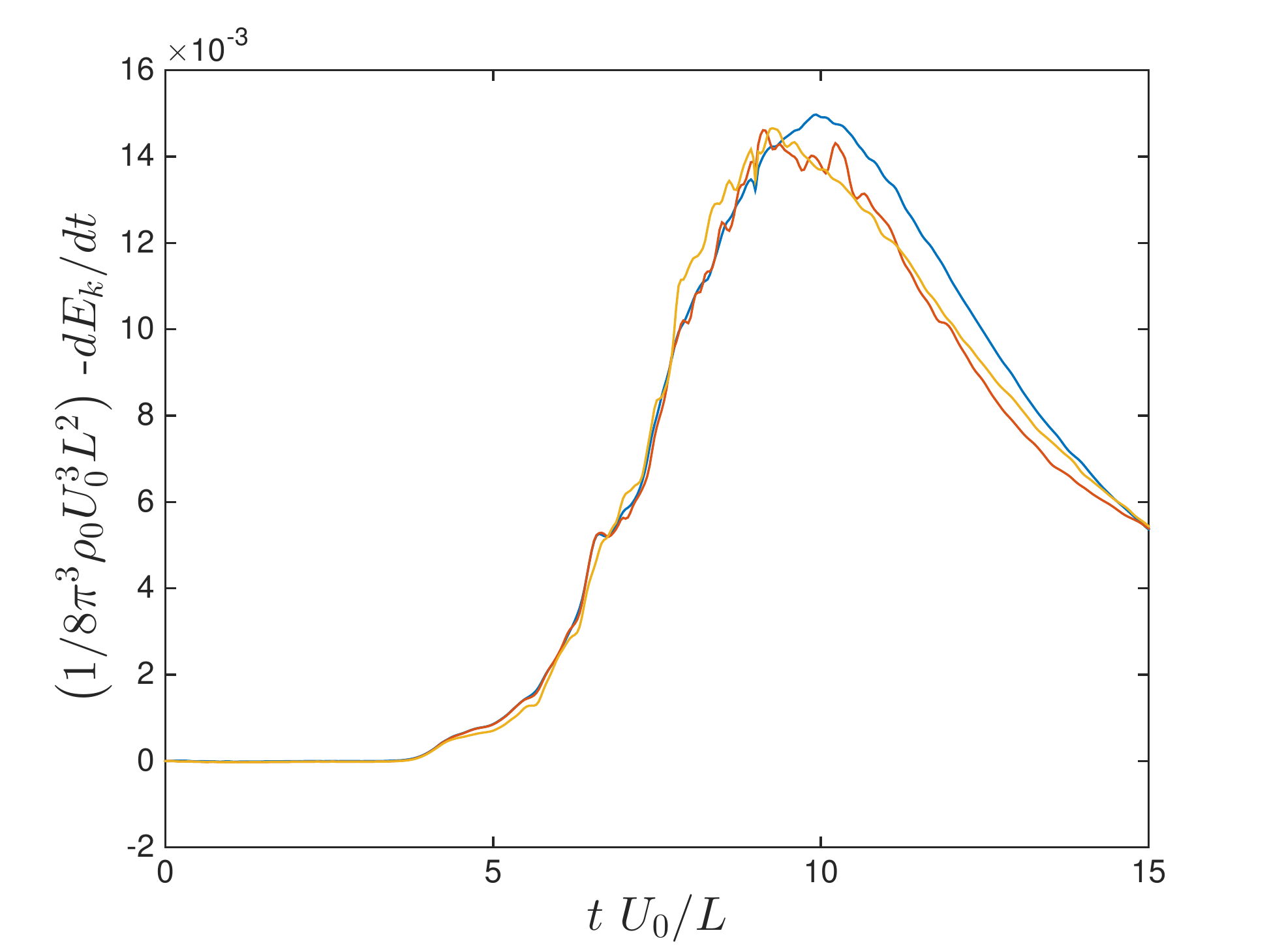}
\caption{\label{dEkdt_Riemann} Time evolution of kinetic energy dissipation rate in the Taylor-Green vortex at $Re = 1600$ (left) and $Re = \infty$ (right) for the Riemann solvers considered.}
\end{figure}

\begin{figure}[t!]
\centering
\includegraphics[width=0.49\textwidth]{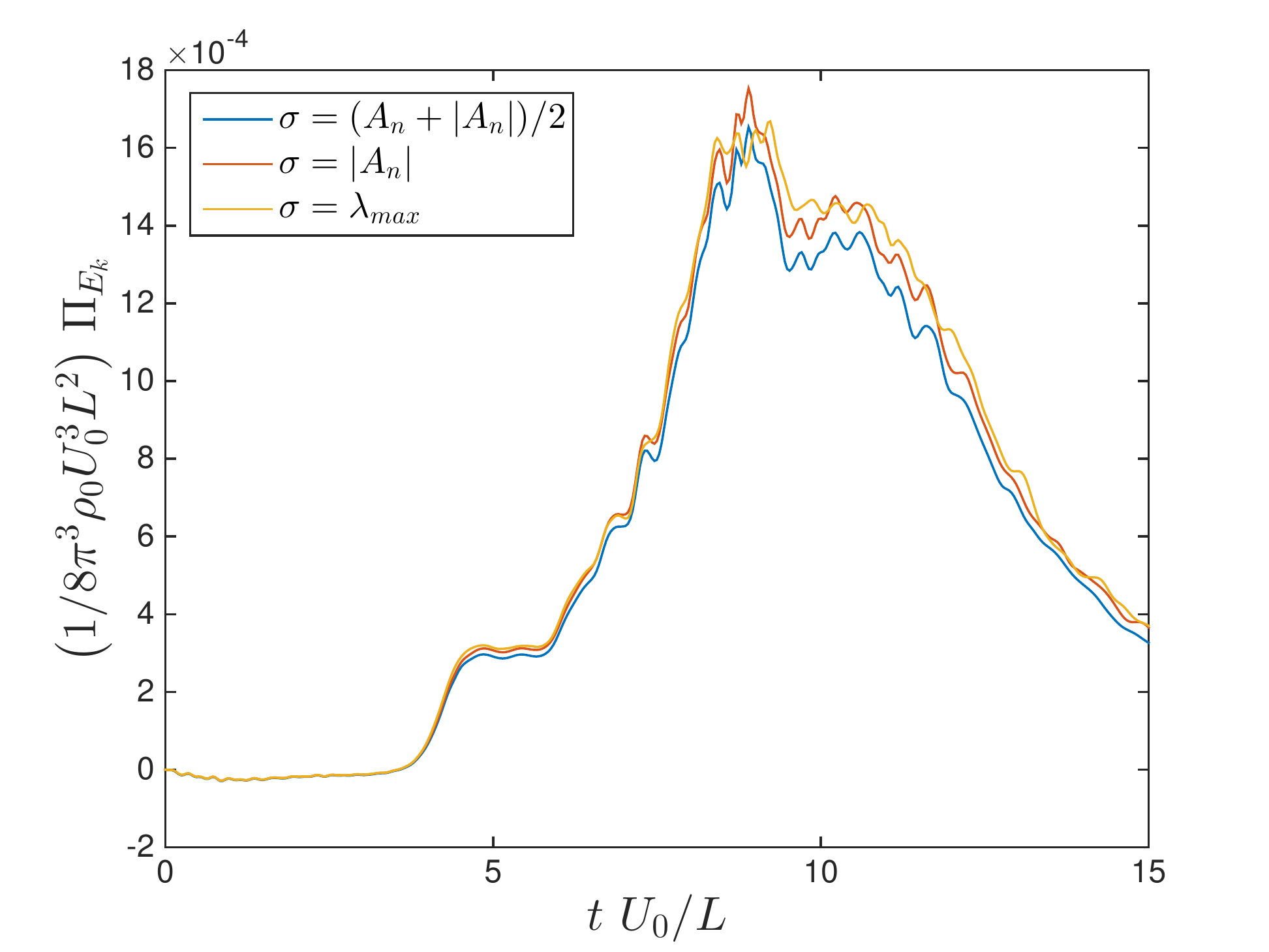}
\includegraphics[width=0.49\textwidth]{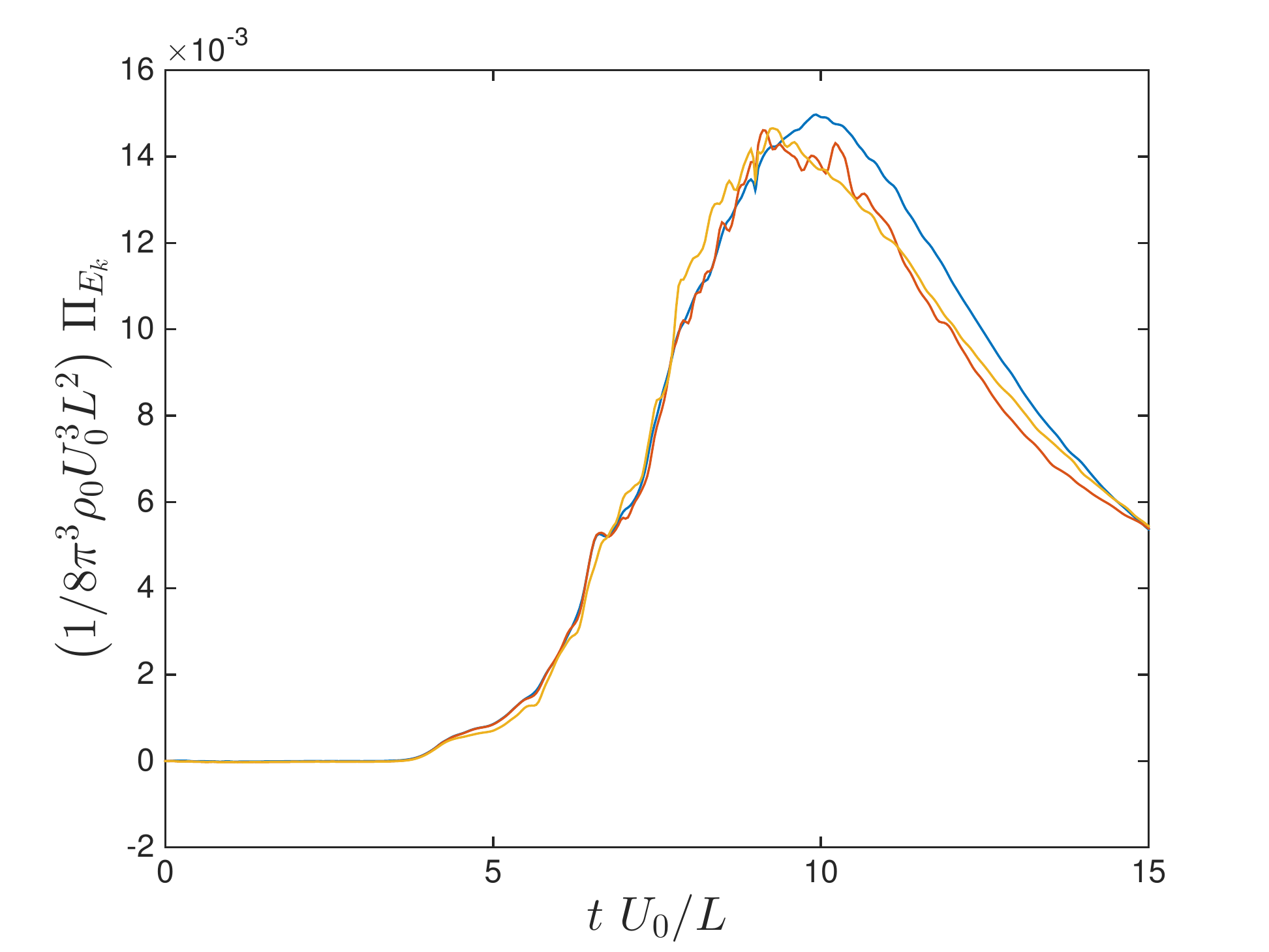}
\caption{\label{numDiss_Riemann} Time evolution of numerical dissipation of kinetic energy, as defined in Eq. \eqref{e:Pi_Ek}, in the Taylor-Green vortex at $Re = 1600$ (left) and $Re = \infty$ (right) for the Riemann solvers considered.}
\end{figure}

Figure \ref{dEkdt_Riemann} shows the time evolution of kinetic energy dissipation rate,
\begin{equation}
\label{e:dEkdt}
\frac{d E_k}{dt} = \frac{d}{dt} \int_{\Omega} \frac{1}{2} \, \rho \norm{\bm{v}}^2 , 
\end{equation}
in the viscous (left) and inviscid (right) Taylor-Green vortex for the three Riemann solvers considered. Figure \ref{numDiss_Riemann} shows the time evolution of the quantity
\begin{equation}
\label{e:Pi_Ek}
\Pi_{E_k} := - \frac{d E_k}{dt} - \int_{\Omega} \mu \norm{\bm{w}}^2 , 
\end{equation}
where $\bm{w} = \nabla \times \bm{v}$ denotes vorticity. 
The second term in the right-hand side of \eqref{e:Pi_Ek} 
corresponds to the dissipation of kinetic energy due to physical mechanisms. 
$\Pi_{E_k}$ is therefore the contribution of the numerical scheme to the dissipation of kinetic energy, and is referred to as the {\it numerical dissipation of kinetic energy}. We note that $\Pi_{E_k}$ should approximately account for the transfer of kinetic energy from resolved scales to subgrid scales\footnote{For the particular case of statistically stationary flows, which is not the case in the Taylor-Green vortex, the transfer of kinetic energy from resolved to subgrid scales is approximately equal to the (more common concept of) viscous dissipation in the subgrid scales.}. 
Also note that Eq. \eqref{e:Pi_Ek} is derived from the incompressible kinetic energy equation since numerical results suggest the incompressible kinetic energy equation is more appropriate than the compressible kinetic energy equation to assess numerical dissipation in nearly incompressible flows \cite{Fernandez:AIAA:17a}.

Tables \ref{t:jumpsRe1600t3}, \ref{t:jumpsRe1600t8}, \ref{t:jumpsReInftyt3} and \ref{t:jumpsReInftyt8} collect the average absolute-value jump across elements on the periodic plane $x = - L \pi$ for each conservation variable $j = 1, ..., 5$, defined as
\begin{equation}
\mathcal{J} ( \bm{u}_{h,j} ) = \frac{ \int_{x = - L \pi} \big| \llbracket \bm{u}_{h,j} \rrbracket_F \big| }{ \int_{x = - L \pi} \big| \langle \bm{u}_{h,j} \rangle_F \big| } , 
\end{equation}
at $Re = 1600$ and $t = 3 \, L / V_0$, $Re = 1600$ and $t = 8 \, L / V_0$, $Re = \infty$ and $t = 3 \, L / V_0$, and $Re = \infty$ and $t = 8 \, L / V_0$, respectively. Here, $\llbracket \bm{u}_{h,j} \rrbracket_F = \bm{u}_{h,j}^{+}|_F - \bm{u}_{h,j}^{-}|_F$ and $\langle \bm{u}_{h,j} \rangle_F = (\bm{u}_{h,j}^{+}|_F + \bm{u}_{h,j}^{-}|_F) / 2$ denote the face jump and face average operators. The term $|_F$ is used to emphasize that the DG solution varies inside each element and is to be evaluated on the element face. 
From these figures and tables, several remarks follow.

({\em i}) The DG scheme adds numerical dissipation when the exact solution contains subgrid scales, that is, after $t \approx 4 \, L / V_0$. This suggests discontinuous Galerkin methods have a built-in (implicit) {\it subgrid-scale model} and introduce additional dissipation in the presence of under-resolved turbulence. 
%
%

({\em ii}) The numerical dissipation is negligible before subgrid scales appear in the flow. Unlike the explicit SGS models in the next section, the implicit model succeeds to detect there are no subgrid scales and does not add numerical dissipation under those conditions.

Observations ({\em i}) and ({\em ii}) imply the numerical dissipation of the DG scheme is positive for under-resolved turbulent flows and vanishes for laminar flows, that is, the implicit model behaves like a dynamical model. This dynamic behavior is justified as follows: On the one hand, if the exact solution does not contain subgrid scales, it is well represented in the DG approximation space and the scheme is in the asymptotic convergence regime. This implies the inter-element jumps are small and in particular $\norm{ \llbracket \bm{u}_{h} \rrbracket } = \mathcal{O}(h^{k+1})$ \cite{Nguyen:12,Peraire:10,Peraire:11}, where $h$ denotes the element size and $k$ is the polynomial order of the DG approximation. Indeed, we note the very small magnitude of the inter-element jumps in Tables \ref{t:jumpsRe1600t3} and \ref{t:jumpsReInftyt3}. Since the amount of numerical dissipation per unit element face area in DG methods is of order $\mathcal{O}(\norm{ \llbracket \bm{u}_{h} \rrbracket }^2)$ \cite{Barth:99,Fernandez:PhD:2018,Fernandez:entrStEuler:2018}, it is therefore negligible when there are no subgrid scales. 
On the other hand, when the exact solution contains subgrid scales (i.e.\ when the simulation becomes under-resolved), the inter-element jumps grow and stabilize the scheme by adding numerical dissipation. Hence, the Riemann solver plays the role of a dynamic SGS model and 
accounts for the effect of the subgrid scales in a 
similar way as explicit models do. We note that, while subgrid scales are most commonly encountered in the simulation of turbulent flows, they may also exist in laminar flows, such as in the Taylor-Green vortex between $t \approx 4 \, L / V_0$ and $t \approx 7-9 \, L / V_0$. 

({\em iii}) Despite under-resolution, no significant differences between Riemann solvers are observed in the viscous case. Even in the inviscid limit the role of the Riemann solver is still moderate. We note that the stabilization matrix $\sigma = \lambda_{max}$ yields much smaller jumps in the momentum fields than the other two stabilization matrices. This is due to the over-upwinding for the momentum equations provided by Lax-Friedrichs-type Riemann solvers at low Mach numbers \cite{Moura:17}, and indicates that DG methods have a (nonlinear) auto-correction mechanism that adapts the magnitude of the inter-element jumps to partially compensate for overshoots in the Riemann solver. This auto-correction mechanism in turn justifies the minor role of the Riemann solver.

({\em iv}) All the Riemann solvers slightly underestimate the peak dissipation with respect to the direct numerical simulation (DNS) data \cite{vanRees:11}. That is, the numerical dissipation is smaller than the true SGS dissipation (only) when the smallest turbulent structures appear. 
Under those conditions, the DG scheme could benefit from the addition of an explicit SGS model. However, as will be discussed in Section \ref{s:TGV_SGS}, this is not the case in practice since the implicit model is partially inhibited by the use of an explicit model.


\begin{table}[t!]
\centering
\begin{tabular}{cccccc}
\hline
 & $\mathcal{J} ( \rho )$ & $\mathcal{J} ( \rho u )$ & $\mathcal{J} ( \rho v )$ & $\mathcal{J} ( \rho w )$ & $\mathcal{J} ( \rho E )$ \\
\hline
\multicolumn{6}{c}{Study of the Riemann solver} \\
\hline
$(A_n + |A_n|)/2$ & $9.8011 \textnormal{E}-6$ &  $2.9010 \textnormal{E}-5$ &  $1.1487   \textnormal{E}-4$ &  $1.7206 \textnormal{E}-4$ &  $1.4890 \textnormal{E}-3$ \\
$|A_n|$ & $8.0801 \textnormal{E}-6$ &  $2.5949 \textnormal{E}-5$ &  $1.0178 \textnormal{E}-4$ &  $1.5361 \textnormal{E}-4$ &  $1.4649 \textnormal{E}-3$  \\
$\lambda_{max}$ & $5.1404 \textnormal{E}-6$ &  $5.7829 \textnormal{E}-6$ &  $2.9427   \textnormal{E}-5$ &  $7.1974 \textnormal{E}-5$ &  $1.2707 \textnormal{E}-3$  \\
\hline
\multicolumn{6}{c}{Study of the SGS model} \\
\hline
ILES & $9.8011 \textnormal{E}-6$ &  $2.9010 \textnormal{E}-5$ &  $1.1487   \textnormal{E}-4$ &  $1.7206 \textnormal{E}-4$ &  $1.4890 \textnormal{E}-3$ \\
Static Smagorinsky & $8.6188 \textnormal{E}-6$ &  $3.3968 \textnormal{E}-5$ &  $1.0798 \textnormal{E}-4$ &  $1.6477 \textnormal{E}-4$ &  $1.4208 \textnormal{E}-3$ \\
Dynamic Smagorinsky & $7.5943 \textnormal{E}-6$ &  $2.7226 \textnormal{E}-5$ &  $9.6104 \textnormal{E}-5$ &  $1.4933 \textnormal{E}-4$ &  $1.4087 \textnormal{E}-3$ \\
Vreman & $8.6570 \textnormal{E}-6$ &  $8.1831 \textnormal{E}-5$ &  $1.1643  \textnormal{E}-4$ &  $1.7149 \textnormal{E}-4$ &  $1.4754 \textnormal{E}-3$ \\
\hline
\end{tabular}
\caption{\label{t:jumpsRe1600t3} Average absolute-value jump across elements on the periodic plane $x = - L \pi$ of the Taylor-Green vortex at $Re = 1600$ and $t = 3 \, L / V_0$.}
\end{table}

\begin{table}[t!]
\centering
\begin{tabular}{cccccc}
\hline
 & $\mathcal{J} ( \rho )$ & $\mathcal{J} ( \rho u )$ & $\mathcal{J} ( \rho v )$ & $\mathcal{J} ( \rho w )$ & $\mathcal{J} ( \rho E )$ \\
\hline
\multicolumn{6}{c}{Study of the Riemann solver} \\
\hline
$(A_n + |A_n|)/2$ & $8.8732 \textnormal{E}-5$  & $2.5180 \textnormal{E}-4$  & $1.3662 \textnormal{E}-3$  & $1.3590 \textnormal{E}-3$  & $2.0834 \textnormal{E}-2$ \\
$|A_n|$ & $8.7631 \textnormal{E}-5$  & $2.0863 \textnormal{E}-4$  & $1.2062 \textnormal{E}-3$  & $1.1662 \textnormal{E}-3$  & $2.0692 \textnormal{E}-2$ \\
$\lambda_{max}$ & $9.0532 \textnormal{E}-5$  & $6.4432 \textnormal{E}-5$  & $3.7080 \textnormal{E}-4$  & $3.6441 \textnormal{E}-4$  & $2.2635 \textnormal{E}-2$ \\
\hline
\multicolumn{6}{c}{Study of the SGS model} \\
\hline
ILES & $8.8732 \textnormal{E}-5$  & $2.5180 \textnormal{E}-4$  & $1.3662 \textnormal{E}-3$  & $1.3590 \textnormal{E}-3$  & $2.0834 \textnormal{E}-2$ \\
Static Smagorinsky & $4.7639 \textnormal{E}-5$  & $1.4676 \textnormal{E}-4$  & $7.4274 \textnormal{E}-4$  & $6.9583 \textnormal{E}-4$  & $1.1127 \textnormal{E}-2$ \\
Dynamic Smagorinsky & $6.4393 \textnormal{E}-5$ & $1.7776 \textnormal{E}-4$ & $8.8756 \textnormal{E}-4$ & $8.6082 \textnormal{E}-4$ & $1.5112 \textnormal{E}-2$ \\
Vreman & $4.9367 \textnormal{E}-5$  & $3.3538 \textnormal{E}-4$  & $8.0182 \textnormal{E}-4$  & $7.4466 \textnormal{E}-4$  & $1.1585 \textnormal{E}-2$ \\
\hline
\end{tabular}
\caption{\label{t:jumpsRe1600t8} Average absolute-value jump across elements on the periodic plane $x = - L \pi$ of the Taylor-Green vortex at $Re = 1600$ and $t = 8 \, L / V_0$.}
\end{table}

\begin{table}[t!]
\centering
\begin{tabular}{cccccc}
\hline
 & $\mathcal{J} ( \rho )$ & $\mathcal{J} ( \rho u )$ & $\mathcal{J} ( \rho v )$ & $\mathcal{J} ( \rho w )$ & $\mathcal{J} ( \rho E )$ \\
\hline
\multicolumn{6}{c}{Study of the Riemann solver} \\
\hline
$(A_n + |A_n|)/2$ & $3.8325 \textnormal{E}-5$ &  $2.5293 \textnormal{E}-4$ &  $1.6103   \textnormal{E}-4$ &  $2.3329 \textnormal{E}-4$ &  $1.9081 \textnormal{E}-3$ \\
$|A_n|$ & $3.0998 \textnormal{E}-5$ &  $2.7685 \textnormal{E}-5$ &  $1.2543 \textnormal{E}-4$ &  $1.8016 \textnormal{E}-4$ &  $1.6249 \textnormal{E}-3$  \\
$\lambda_{max}$ & $7.7441 \textnormal{E}-6$ &  $5.7671 \textnormal{E}-6$ &  $3.4419 \textnormal{E}-5$ &  $7.3344 \textnormal{E}-5$ &  $1.5169 \textnormal{E}-3$  \\
\hline
\multicolumn{6}{c}{Study of the SGS model} \\
\hline
ILES & $3.8325 \textnormal{E}-5$ &  $2.5293 \textnormal{E}-4$ &  $1.6103   \textnormal{E}-4$ &  $2.3329 \textnormal{E}-4$ &  $1.9081 \textnormal{E}-3$ \\
Static Smagorinsky & $1.9543 \textnormal{E}-5$ &  $3.1352 \textnormal{E}-5$ &  $1.2723 \textnormal{E}-4$ &  $1.8851 \textnormal{E}-4$ &  $1.6236 \textnormal{E}-3$ \\
Dynamic Smagorinsky & $2.9271 \textnormal{E}-5$ &  $2.7791 \textnormal{E}-5$ &  $1.1150 \textnormal{E}-4$ &  $1.7016 \textnormal{E}-4$ &  $1.5298 \textnormal{E}-3$ \\
Vreman & $1.7323 \textnormal{E}-5$ &  $3.5460 \textnormal{E}-5$ &  $1.3224 \textnormal{E}-4$ &  $1.9255 \textnormal{E}-4$ &  $1.6687 \textnormal{E}-3$ \\
\hline
\end{tabular}
\caption{\label{t:jumpsReInftyt3} Average absolute-value jump across elements on the periodic plane $x = - L \pi$ of the Taylor-Green vortex at $Re = \infty$ and $t = 3 \, L / V_0$.}
\end{table}

\begin{table}[t!]
\centering
\begin{tabular}{cccccc}
\hline
 & $\mathcal{J} ( \rho )$ & $\mathcal{J} ( \rho u )$ & $\mathcal{J} ( \rho v )$ & $\mathcal{J} ( \rho w )$ & $\mathcal{J} ( \rho E )$ \\
\hline
\multicolumn{6}{c}{Study of the Riemann solver} \\
\hline
$(A_n + |A_n|)/2$ & $2.3608 \textnormal{E}-4$ &  $1.2680 \textnormal{E}-2$ &  $4.7186 \textnormal{E}-3$ &  $4.9416 \textnormal{E}-3$ &  $5.3469 \textnormal{E}-2$ \\
$|A_n|$ & $2.4658 \textnormal{E}-4$ &  $1.0320 \textnormal{E}-3$ &  $4.7063 \textnormal{E}-3$ &  $4.7594 \textnormal{E}-3$ &  $5.8251 \textnormal{E}-2$  \\
$\lambda_{max}$ & $4.0504 \textnormal{E}-4$ &  $5.6643 \textnormal{E}-4$ &  $1.8876 \textnormal{E}-3$ &  $1.9245 \textnormal{E}-3$ &  $1.0114 \textnormal{E}-1$  \\
\hline
\multicolumn{6}{c}{Study of the SGS model} \\
\hline
ILES & $2.3608 \textnormal{E}-4$ &  $1.2680 \textnormal{E}-2$ &  $4.7186 \textnormal{E}-3$ &  $4.9416 \textnormal{E}-3$ &  $5.3469 \textnormal{E}-2$ \\
Static Smagorinsky & $1.1676 \textnormal{E}-4$ &  $3.7351 \textnormal{E}-4$ &  $1.8456 \textnormal{E}-3$ &  $2.1074 \textnormal{E}-3$ &  $2.7445 \textnormal{E}-2$ \\
Dynamic Smagorinsky & $1.7943 \textnormal{E}-4$ &  $1.2174 \textnormal{E}-3$ &  $2.7746 \textnormal{E}-3$ &  $3.0035 \textnormal{E}-3$ &  $4.3530 \textnormal{E}-2$ \\
Vreman & $1.3084 \textnormal{E}-4$ &  $4.0653 \textnormal{E}-4$ &  $2.1125 \textnormal{E}-3$ &  $2.3270 \textnormal{E}-3$ &  $3.1185 \textnormal{E}-2$ \\
\hline
\end{tabular}
\caption{\label{t:jumpsReInftyt8} Average absolute-value jump across elements on the periodic plane $x = - L \pi$ of the Taylor-Green vortex at $Re = \infty$ and $t = 8 \, L / V_0$.}
\end{table}

\begin{figure}
 \centering
 {\includegraphics[width=0.49\textwidth]{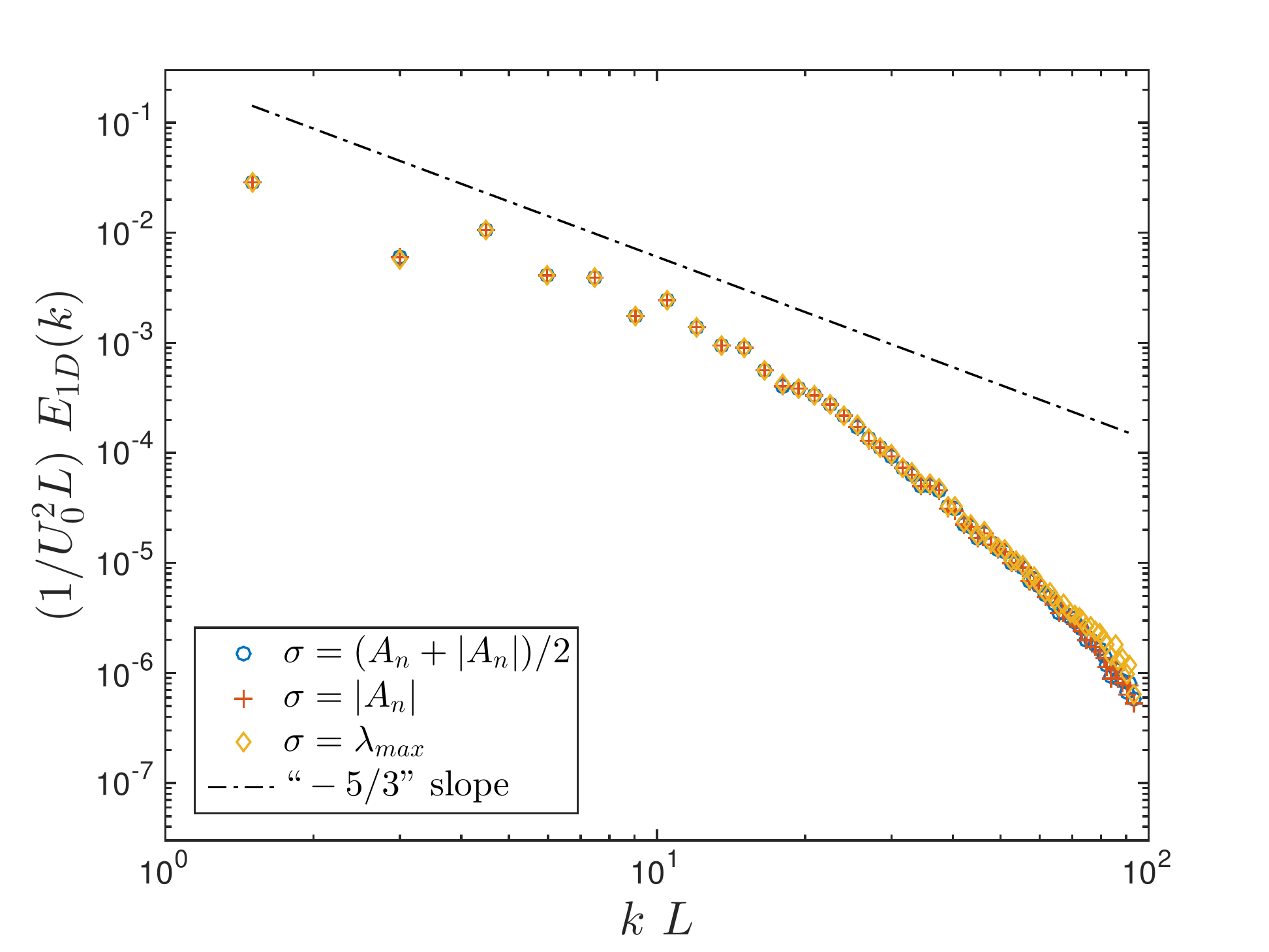}}
  \hfill {\includegraphics[width=0.49\textwidth]{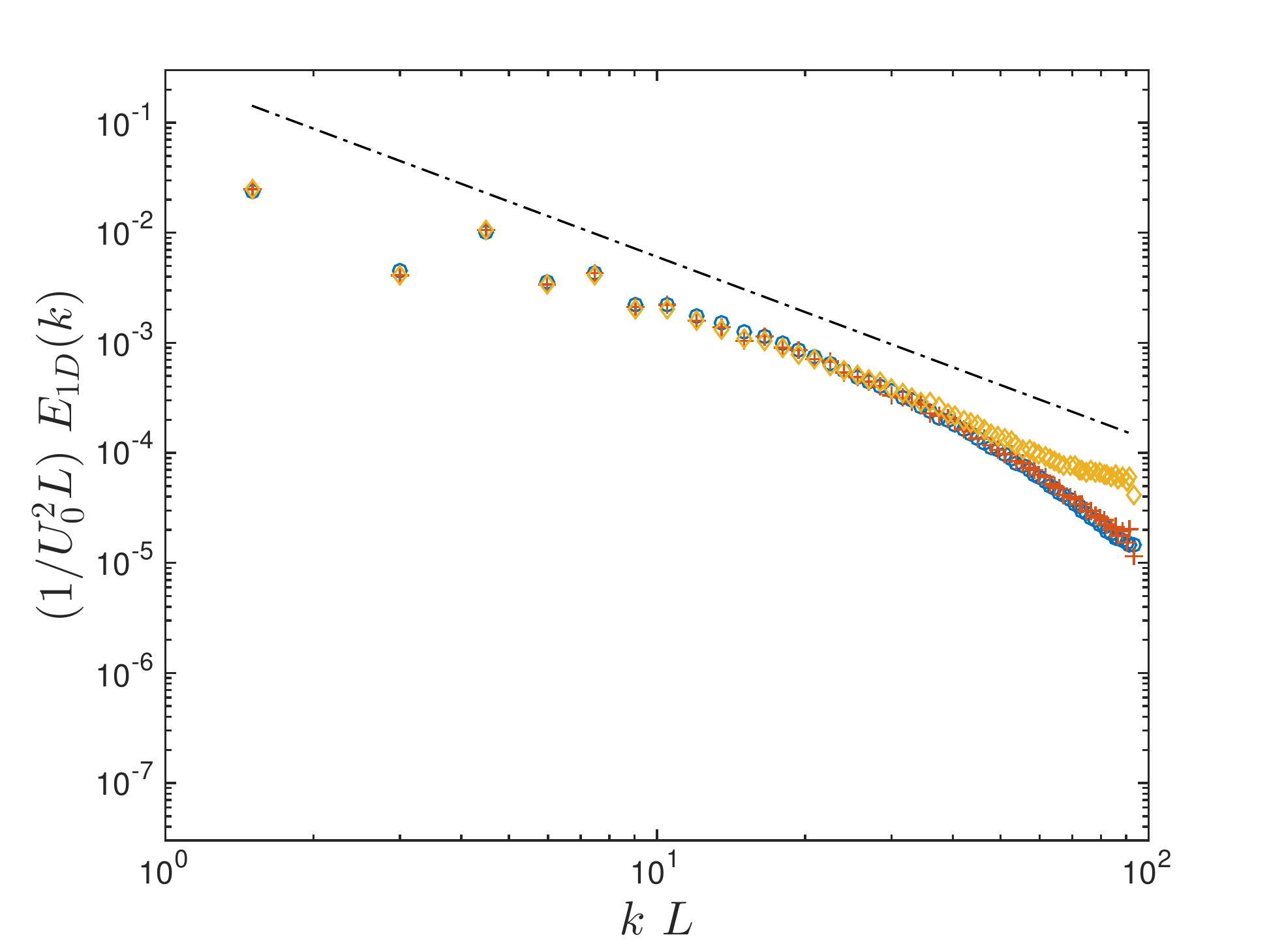}}
  \caption{One-dimensional kinetic energy spectrum in the Taylor-Green vortex at $Re = 1600$ and $t = 8 \, L / V_0$ (left), and $Re = \infty$ and $t = 9 \, L / V_0$ (right) for the Riemann solvers considered.}\label{TGV_kinEnSpec_Riemann}
 \end{figure}

Figure \ref{TGV_kinEnSpec_Riemann} shows the one-dimensional kinetic energy spectra at $t = 8 \, L / V_0$ and $t = 9 \, L / V_0$ for the viscous and inviscid Taylor-Green vortex, respectively. In all cases, we observe an inertial range in which the spectrum follows a power law with exponent close to the theoretical value of $-5/3$ \cite{Kolmogorov:41}. The inertial range is followed by a dissipative range, until the grid Nyquist wavenumber $k_{N} = 96.0 / L$ is achieved and no smaller scales exist in the discretization. In the viscous case, the inertial range extends up to $k \approx 20 / L$, whereas DNS results \cite{vanRees:11} indicate it extends until $k \approx 40 / L$. This numerically-induced premature end of the inertial range is predicted by eigenanalysis \cite{Moura:15a}, Fourier analysis \cite{Alhawwary:2018} and non-modal analysis \cite{Fernandez:nonModal:2018} theory. 
Also, note that no significant differences in the spectra are observed between Riemann solvers in the viscous case (neither at the times shown nor at other times); which shows that at this Reynolds number the viscous dissipation has a much larger impact on the dynamics of all the scales than the Riemann solver. 
We also note the presence of an energy {\it pileup} at large wavenumbers in the inviscid case for the Riemann solver $\sigma = \lambda_{max}$. These pre-dissipative {\it bumps} are predicted by eigenanalysis \cite{Moura:15a} and non-modal analysis \cite{Fernandez:nonModal:2018} theory, and are consistent with results in the literature for inviscid low Mach numbers flows when using Riemann solvers that are based on the maximum-magnitude eigenvalue of $\bm{A}_n$ \cite{Moura:16,Moura:17}. 
From these results and the insights from eigenanalysis and non-modal analysis, the Riemann solver has some impact on the dynamics of the smallest resolved scales, particularly at low Mach numbers and high Reynolds numbers (more precisely, at low Mach numbers and high cell P\'eclet numbers); whereas the large-scale dynamics are affected to a much lesser extent by the Riemann solver. As the Mach number increases and the cell P\'eclet number decreases, the choice of Riemann solver is expected to have a smaller impact on the behavior of the small scales.


\subsubsection{\label{s:TGV_SGS}Subgrid-scale model study}

\begin{figure}
\centering
\includegraphics[width=0.49\textwidth]{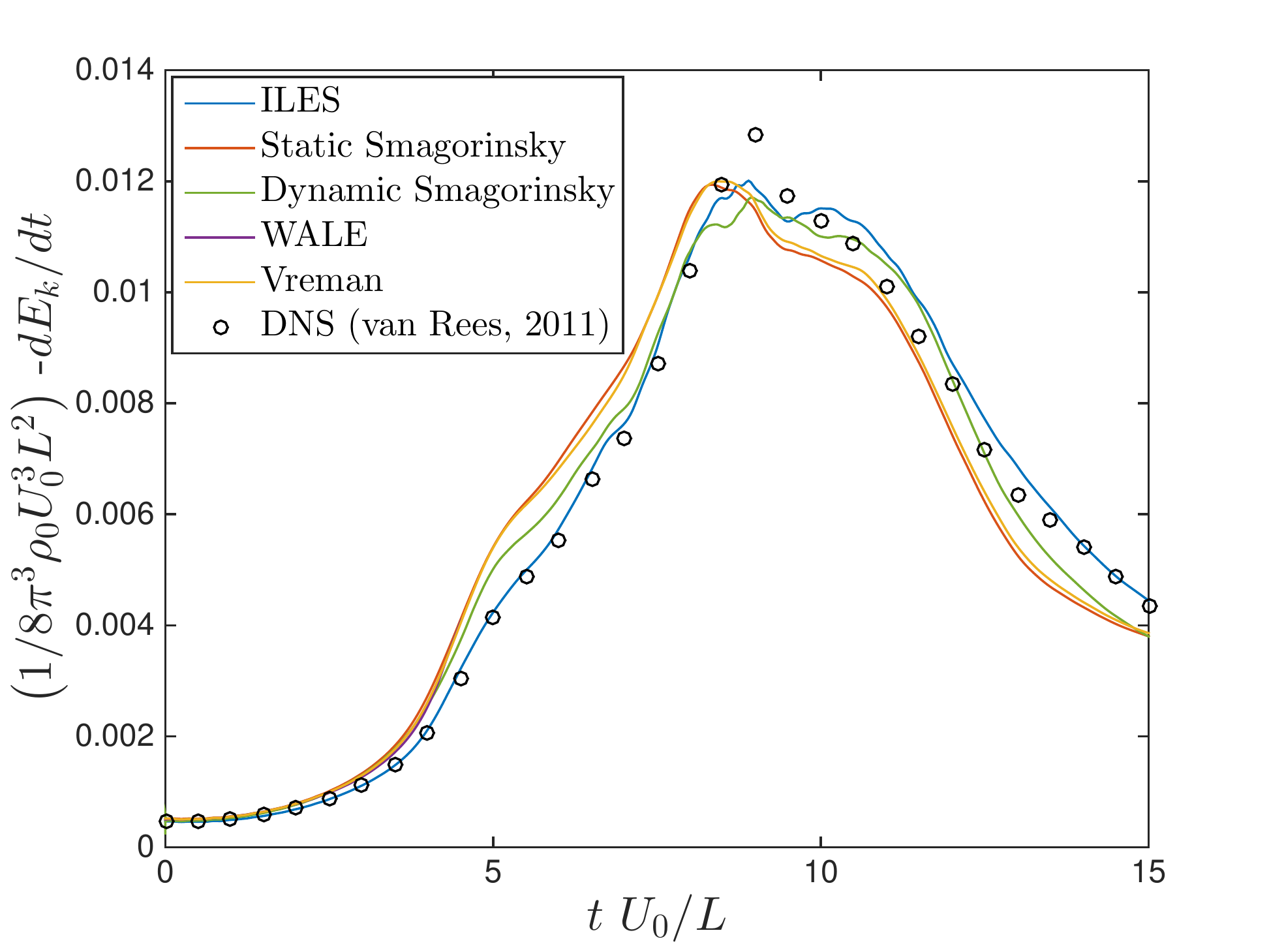}
\includegraphics[width=0.49\textwidth]{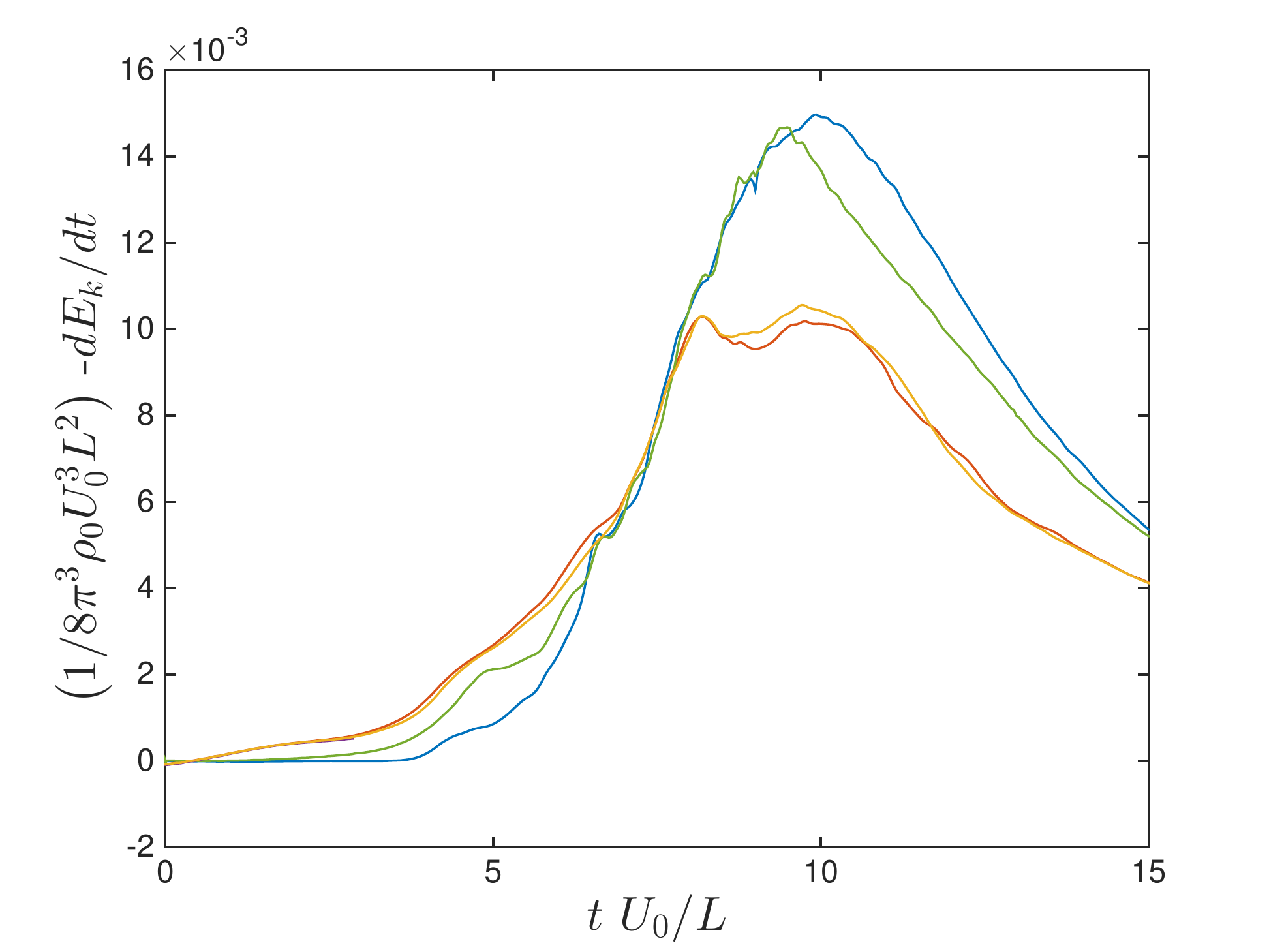}
\caption{\label{dEkdt_SGS} Time evolution of kinetic energy dissipation rate in the Taylor-Green vortex at $Re = 1600$ (left) and $Re = \infty$ (right) for ILES, static Smagorinsky, dynamic Smagorinsky, WALE and Vreman LES.}
\end{figure}

\begin{figure}
\centering
\includegraphics[width=0.49\textwidth]{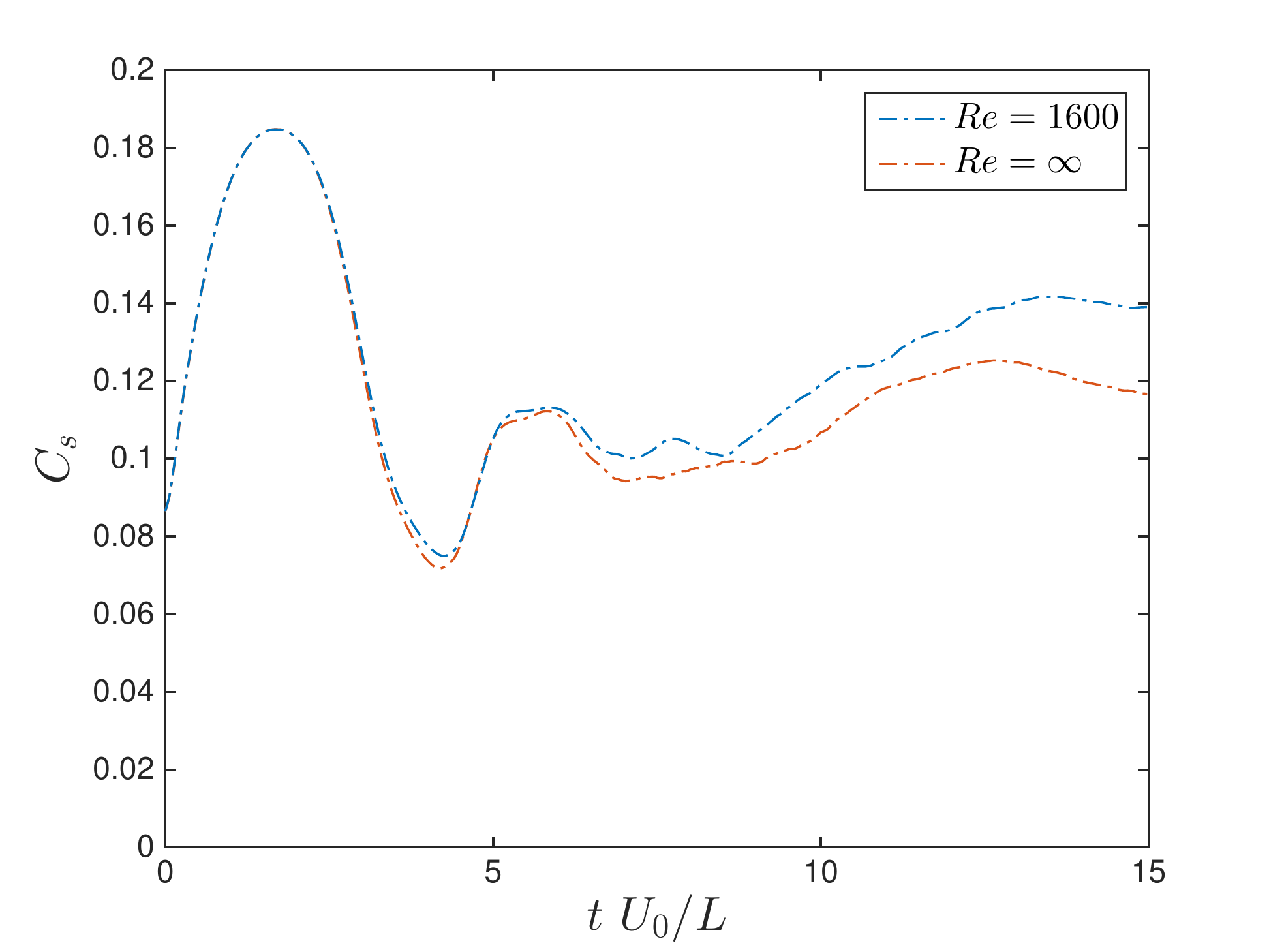}
\caption{\label{f:evolutionCs_dynSmag_TGV} Time evolution of volume-averaged dynamic Smagorinsky constant in the Taylor-Green vortex. Note $C_s = 0.16$ in static Smagorinsky \cite{Gatski:09}.}
\end{figure}

\begin{figure}[t!]
\centering
{\includegraphics[width=0.24\textwidth]{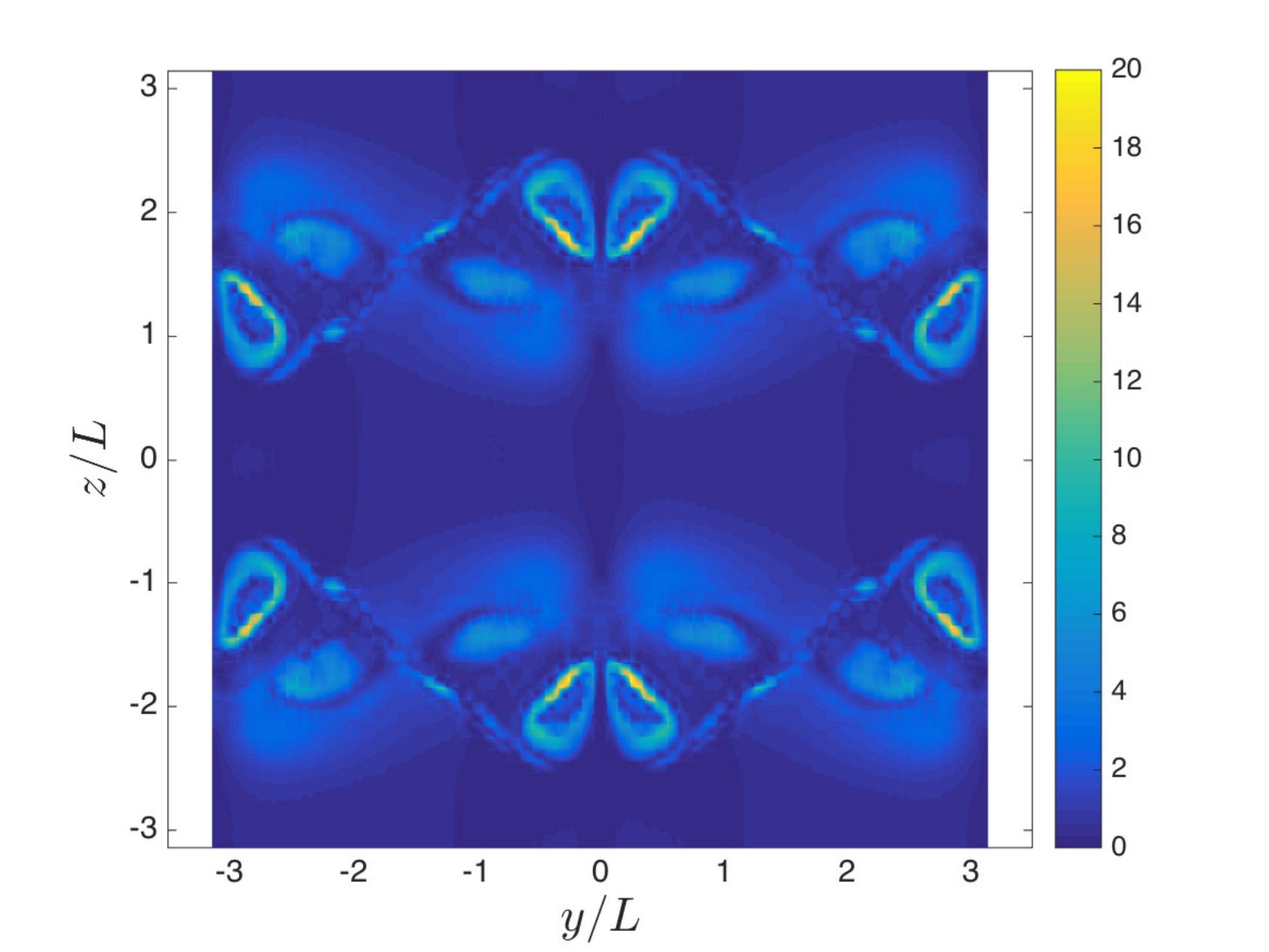}}
\hfill {\includegraphics[width=0.24\textwidth]{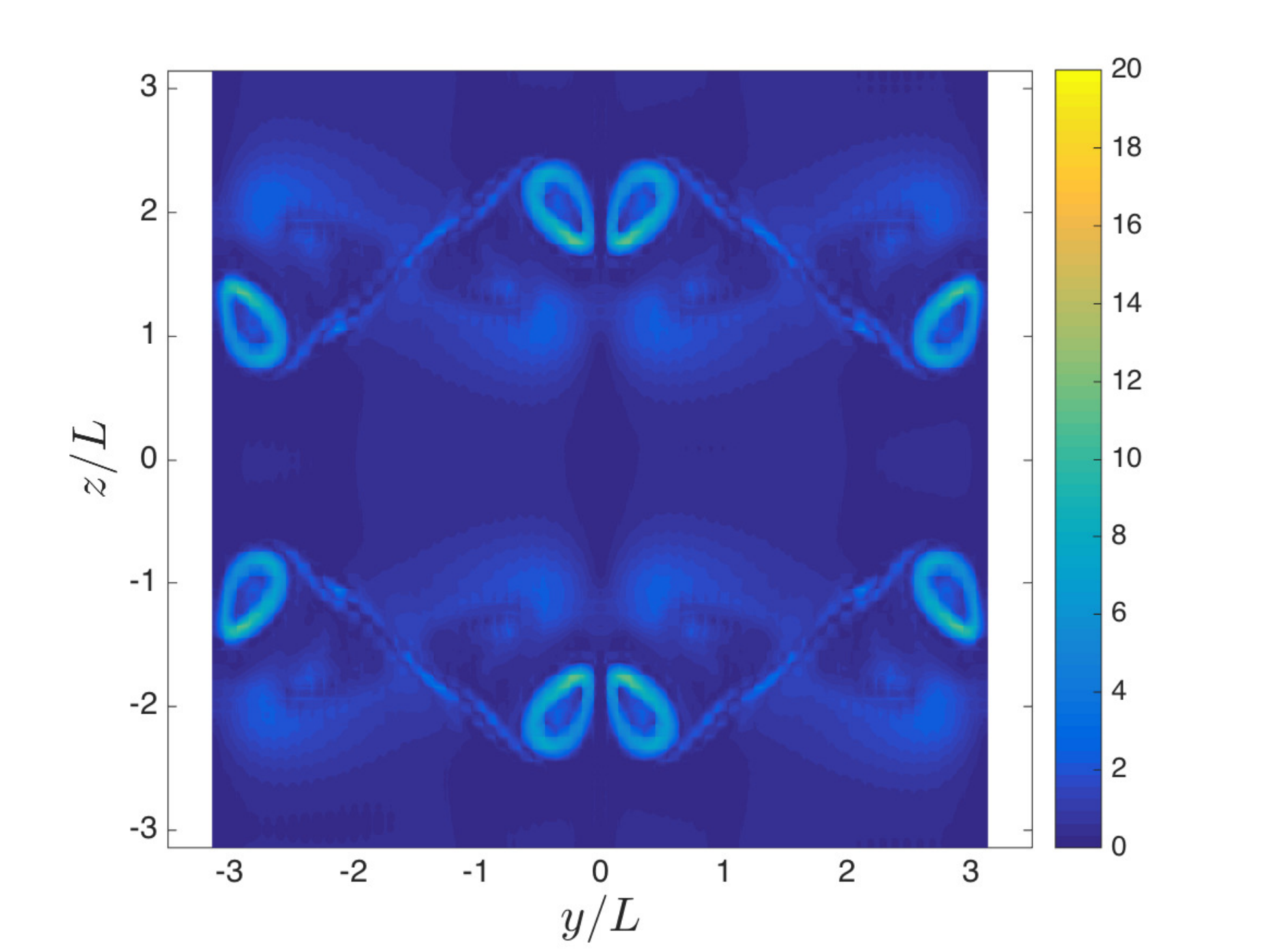}}
\hfill {\includegraphics[width=0.24\textwidth]{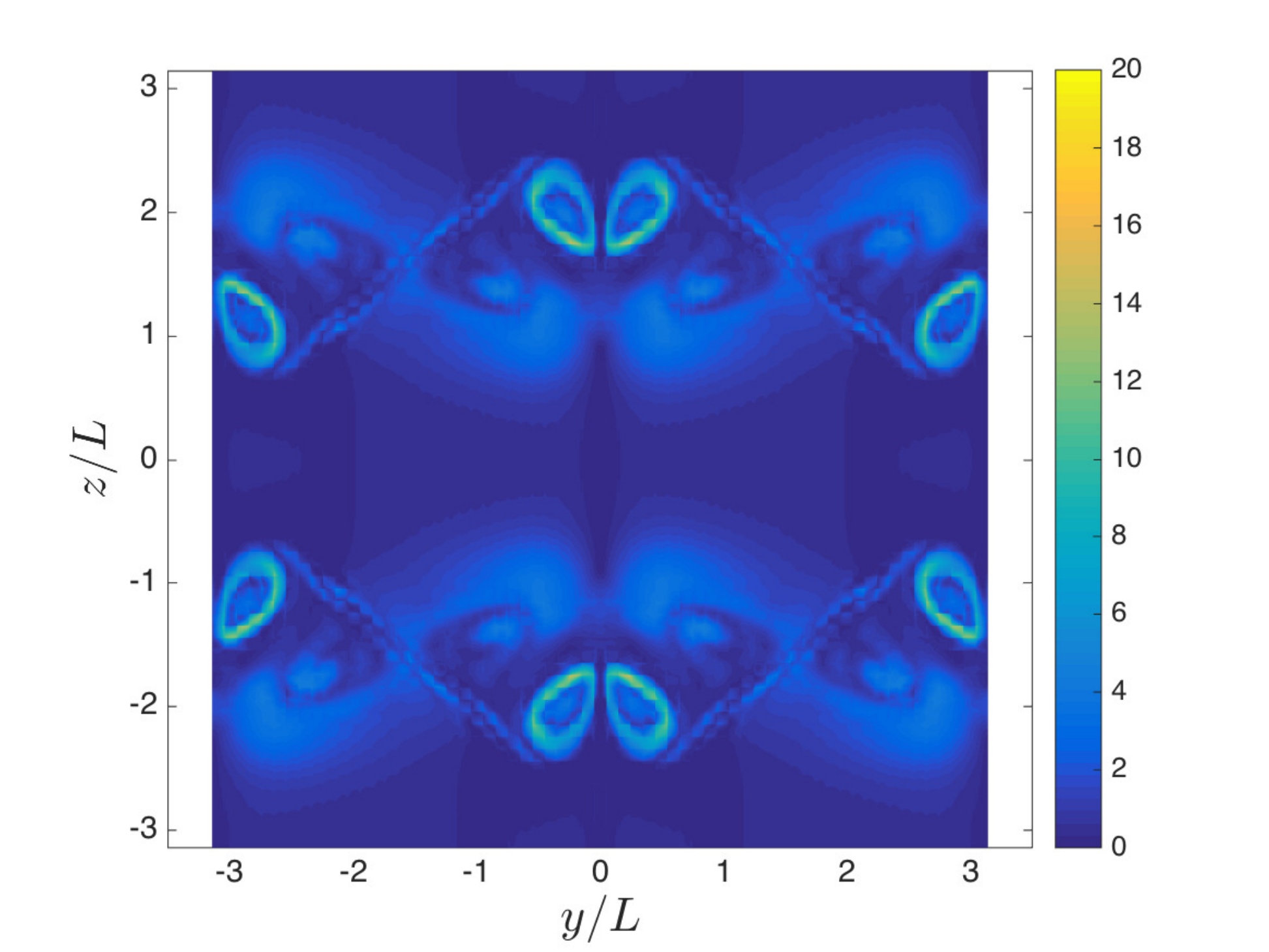}}
\hfill {\includegraphics[width=0.24\textwidth]{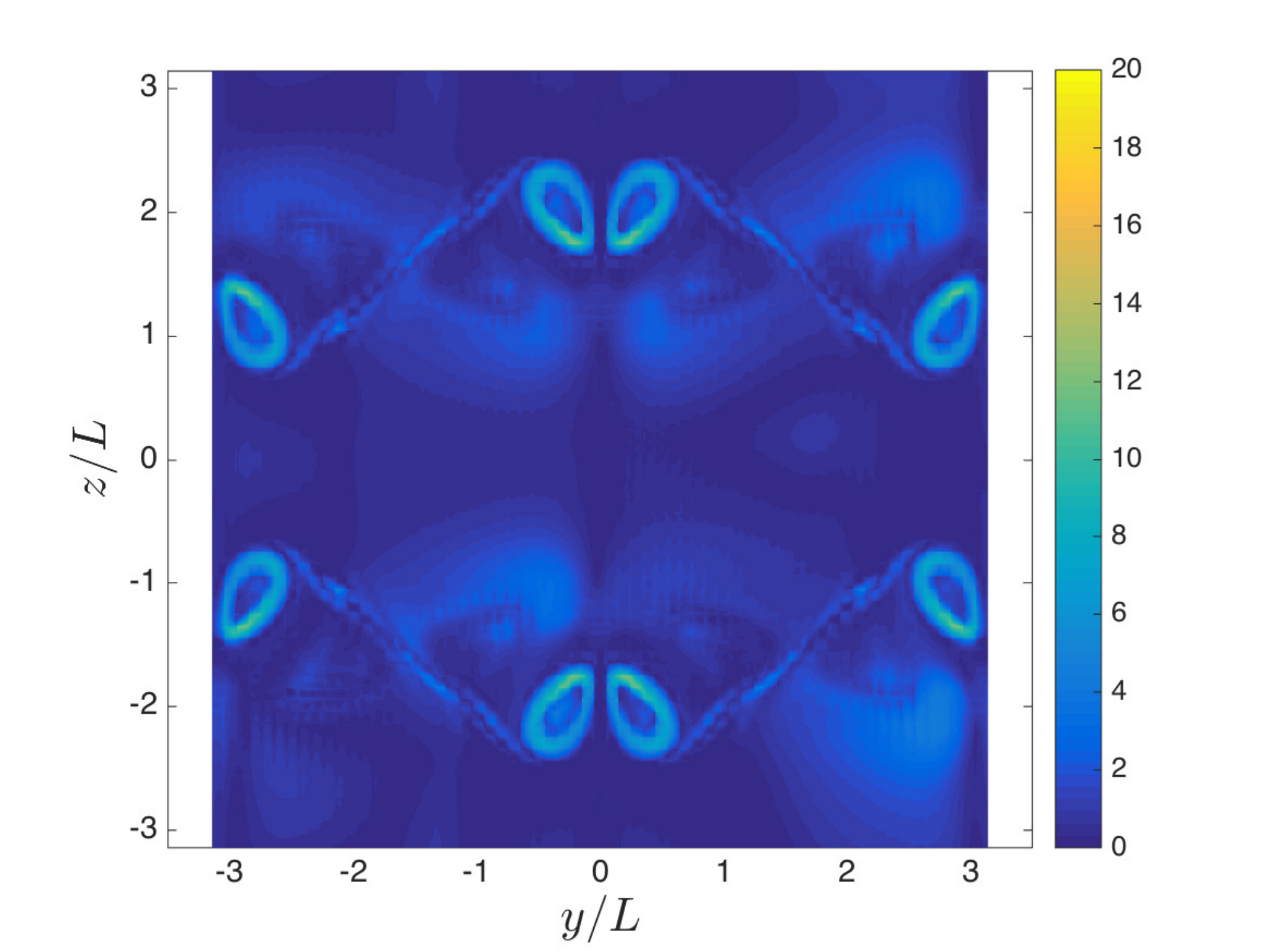}}
\caption{Snapshot of vorticity norm $\norm{\bm{\omega}} L / V_0$ on the periodic plane $x = - L \pi$ of the viscous Taylor-Green vortex at $t = 8 \, L / V_0$. Left to right: ILES, static Smagorinsky, dynamic Smagorinsky and Vreman. The WALE model led to nonlinear instability and the simulation breakdown at $t \approx 4.59 \, L / V_0$.}\label{vortMag_t8}
\end{figure}

Figure \ref{dEkdt_SGS} shows the time evolution of kinetic energy dissipation rate with ILES, the static Smagorinsky, dynamic Smagorinsky, WALE and Vreman models at both Reynolds numbers. The time evolution of the volume-averaged dynamic Smagorinsky constant is shown in Figure \ref{f:evolutionCs_dynSmag_TGV}. The average absolute-value jump across elements on the periodic plane $x = - L \pi$ for the viscous and inviscid cases, both at $t = 3 \, L / V_0$ and $t = 8 \, L / V_0$, are collected in Tables \ref{t:jumpsRe1600t3}, \ref{t:jumpsRe1600t8}, \ref{t:jumpsReInftyt3} and \ref{t:jumpsReInftyt8}. Also, Figure \ref{vortMag_t8} shows snapshots of the vorticity norm on $x = - L \pi$ for the viscous case at $t = 8 \, L / V_0$. 
We recall that the stabilization matrix \eqref{e:stabMatrices1} is used for the SGS study. These results can be summarized as follows:

({\em i}) The eddy viscosity in all the explicit SGS models fails to vanish when there are no subgrid scales in the flow, and produces unphysical dissipation during this phase and also during the under-resolved laminar phase. Among the explicit models, dynamic Smagorinsky introduces the least amount of dissipation in these two phases. We note that a larger dissipation of kinetic energy corresponds to effectively solving a lower Reynolds number flow; which is consistent with the maximum dissipation rate occuring at an earlier time that is characteristic of the Taylor-Green vortex at lower Reynolds numbers \cite{Brachet:91}.

We emphasize that the built-in stabilization due to inter-element jumps (i.e.\ the implicit subgrid-scale model) in the DG scheme provides a more accurate mechanism to detect the absence of subgrid scales, in which case very little dissipation is added, than the explicit models. This is a critical advantage of the implicit model to simulate transitional flows.

({\em ii}) When subgrid scales appear in the flow, the amount of dissipation introduced by the implicit model is closer to the true SGS value than that introduced by the explicit models. Only in the inviscid case in fully turbulent regime, the dynamic Smagorinsky model performs similarly to the implicit model. 
We note that, whenever the simulation is under-resolved, the inter-element jumps with an explicit SGS model are much smaller than with no model (see Tables \ref{t:jumpsRe1600t3}$-$\ref{t:jumpsReInftyt8}); which is due to the additional stabilization provided by the eddy viscosity. The use of an explicit model therefore partially inhibits the implicit model, and 
in fact the dissipation of kinetic energy in the turbulent regime is smaller with the explicit models.

({\em iii}) While the inter-element jumps in the numerical solution are smaller and the vorticity norm field is smoother with an explicit SGS model (due to the eddy viscosity dissipation), ILES provides more accurate results. Therefore, lack of smoothness in the DG solution is not an indicator for low solution quality --and, as discussed previously, the inter-element jumps are actually responsible for the built-in model in the scheme.

({\em iv}) Regarding the relative performance of the explicit models, no major differences are observed between static Smagorinsky and Vreman. The WALE model led to nonlinear instability and the simulation breakdown at $t \approx 4.59 \, L / V_0$ and $2.75 \, L / V_0$ in the viscous and inviscid cases, respectively. This lack of robustness is due to the high nonlinearity of the WALE model and may limit its applicability with high-order DG methods. 
The dynamic Smagorinsky model provides the most accurate representation of the subgrid scales among the explicit models.

\begin{figure}
 \centering
 {\includegraphics[width=0.49\textwidth]{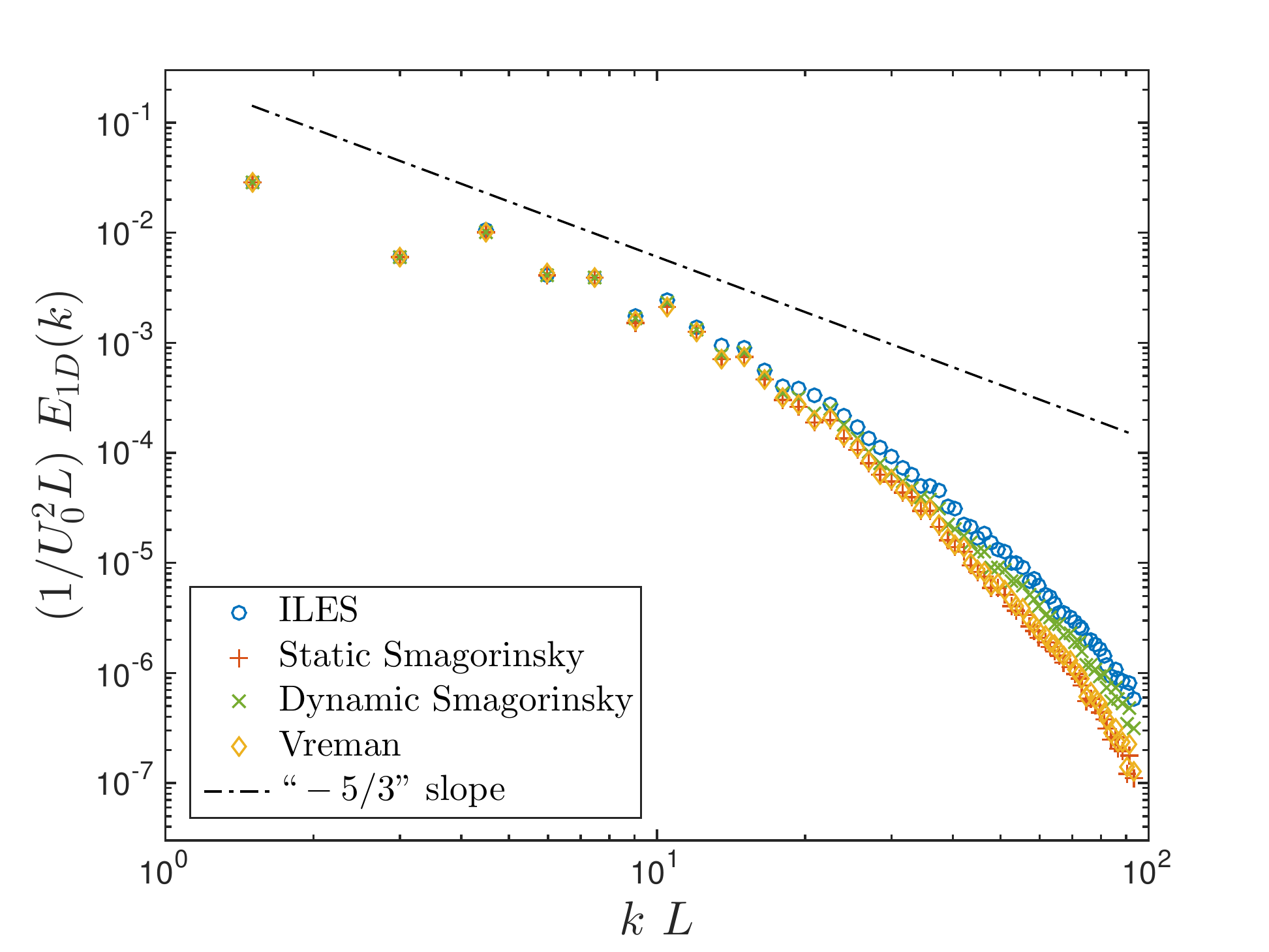}}
  \hfill {\includegraphics[width=0.49\textwidth]{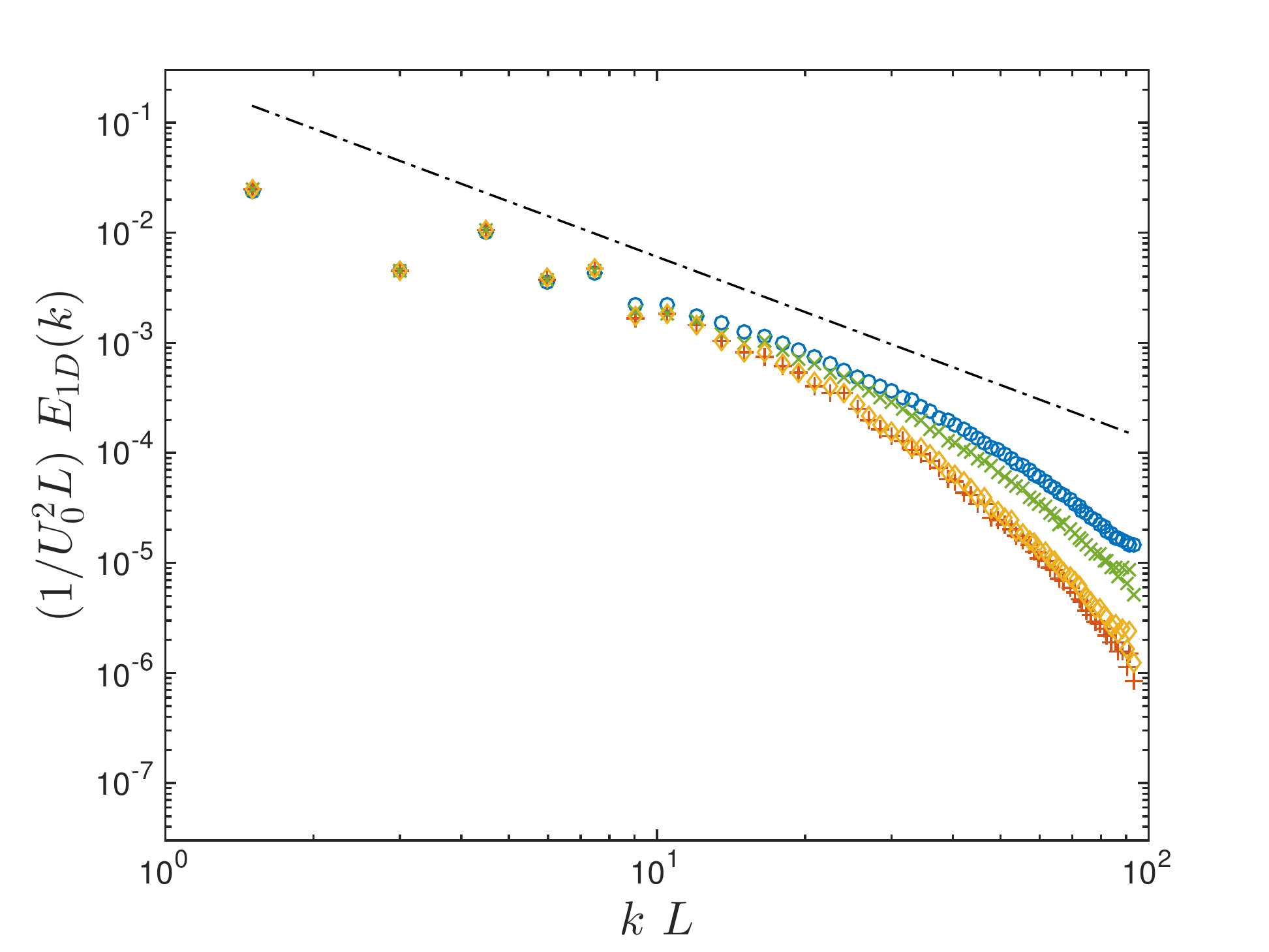}}
  \caption{One-dimensional kinetic energy spectrum in the Taylor-Green vortex at $Re = 1600$ and $t = 8 \, L / V_0$ (left), and $Re = \infty$ and $t = 9 \, L / V_0$ (right) for ILES, static Smagorinsky, dynamic Smagorinsky and Vreman. The WALE model led to nonlinear instability and the simulation breakdown at $t \approx 4.59 \, L / V_0$ and $2.75 \, L / V_0$ in the viscous and inviscid cases, respectively.}\label{TGV_kinEnSpec_SGS}
 \end{figure}

Figure \ref{TGV_kinEnSpec_SGS} shows the one-dimensional kinetic energy spectra for ILES, the static Smagorinsky, dynamic Smagorinsky and Vreman models at $t = 8 \, L / V_0$ and $t = 9 \, L / V_0$ for the viscous and inviscid cases, respectively. The WALE model crashed before these times. We recall that the grid Nyquist wavenumber is $k_{N} = 96.0 / L$. The kinetic energy spectrum in ILES agrees with the theoretical $-5/3$ slope of decay of the inertial range for a larger range of wavenumbers than the explicit models. Small differences are observed between the static Smagorinsky, WALE and Vreman models, both at the times shown as well as at all other times. The dynamic Smagorinsky spectrum is in between that of the implicit LES and those of the static models. All the explicit models, especially the static ones, dissipate kinetic energy at larger scales than the implicit model, and in particular at scales that are much larger than the grid Nyquist wavenumber. This is consistent with eigenanalysis \cite{Ainsworth:2004,Moura:15a}, Fourier analysis \cite{Alhawwary:2018} and non-modal analysis \cite{Fernandez:nonModal:2018}.  
The fact that explicit models dissipate energy at larger scales than the implicit model may have important consequences in practice, as discussed in Sections \ref{s:channel} and \ref{s:conclusions}.

\subsection{Summary}

\noindent The Taylor-Green vortex results indicate that:

\begin{itemize}
\item Discontinuous Galerkin methods 
have a built-in (implicit) subgrid-scale model and numerically dissipate kinetic energy in under-resolved turbulence simulations. 
The inter-element jumps and the Riemann solver are responsible for the implicit model.
\item The amount of dissipation implicitly introduced by the DG scheme is closer to the actual dissipation in the subgrid scales than that with the explicit models considered. 
\item The implicit dissipation is more localized near the grid Nyquist wavenumber (i.e.\ more localized in the smallest resolved scales) than that introduced by the explicit models.
\item The implicit model does not add dissipation when there are no subgrid scales, whereas the explicit models do. That is, the implicit model behaves like a dynamic model.
\end{itemize}
Except for the second remark above, these numerical observations have been justified by 
theoretical studies of DG methods. The second observation could be explained by relating the DG stabilization due to the inter-element jumps with the subgrid-scale closure terms arising from variational multiscale (VMS) \cite{Hughes:1998} or Mori-Zwanzig (MZ) \cite{Parish:2017,Parish:2017b} approaches. 
For one-dimensional linear convection, the MZ-VMS procedure with the assumptions of finite memory and linear quadrature, referred to as $\tau$-MZ-VMS, actually leads to a subgrid-scale closure term that is equivalent to the standard upwind flux \cite{Parish:MZVMS:2018}, i.e.\ the implicit model with standard upwinding is the same as that given by $\tau$-MZ-VMS. 
This analogy is less straightforward in the case of the Navier-Stokes equations and has not been investigated in this paper.


\section{\label{s:channel}Turbulent channel flow}

\subsection{\label{s:channel_caseDescription}Case description}

We consider the turbulent channel flow \cite{LM:2015} at $Re_{\tau} = 182$ and $544$, where $Re_{\tau} = \rho_0 u_{\tau} \delta / \mu$ is the Reynolds number based on the volume-averaged density $\rho_0$, the friction velocity $u_{\tau} = \sqrt{\tau_w / \rho_0}$ and the channel half-width $\delta$, and where $\tau_w$ denotes the mean wall shear stress. The bulk Mach number is $M_b = U_b / c = 0.2$, where $U_b = \int_{0}^{2 \delta} {u(y) \, dy} \, / \, 2 \delta$ and $c$ are the bulk velocity and the speed of sound at the mean temperature. 
The flow is statistically stationary and driven by a uniform pressure gradient, which varies in time to ensure that the mass flux through the channel remains constant. The top and bottom walls of the channel are no-slip and adiabatic. This completes the non-dimensional description of the problem.

As is customary in channel flows, the three velocity components are denoted by $(u,v,w)$, the time-averaged velocity by a capital letter, the fluctuations by a prime, and the ensemble average by $\langle \, \cdot \, \rangle$. Thus, $U = \langle u \rangle$ and $u = U + u'$, and similarly for the wall-normal and spanwise velocity components. 

\subsection{Details of the numerical discretization}

The channel flow is simulated in a doubly-periodic domain $\Omega = [0 , 4 \pi \delta ) \times [0 , 2 \delta) \times [0 , 2 \pi \delta)$. 
Periodicity is imposed along the $x$ (streamwise) and $z$ (spanwise) directions. The computational domain is partitioned into a $48 \times 32 \times 40$ Cartesian grid and the third-order Interior Embedded DG (IEDG) scheme \cite{Fernandez:16a} is used for the spatial discretization. The element size is constant along the streamwise and spanwise directions. The high-order nodes in the wall-normal direction $y$ are uniformly distributed in a mapped coordinate $\xi$ that is related to $y$ through
\begin{equation}
\label{e:yNodesDistribution}
\frac{y}{\delta} = \frac{\sin (\xi \pi / 2) }{\sin (\pi / 2) } + 1 , \quad -1 \leq \xi \leq 1 . 
\end{equation}
The distance between high-order nodes (in wall units) is summarized in Table \ref{infoChannelDiscr}. As is customary, the near-wall velocity, time and length scales for non-dimensionalization are $u_{\tau}$, $\mu / \rho_0 u_{\tau}^2$ and $\mu / \rho_0 u_{\tau}$, respectively, and the superscript $+$ is used to indicate that a quantity is expressed in wall units. 
A run-up time $T_1^+ = 2000$ is used for the flow to achieve its stationary distribution on the chaotic attractor. The flow statistics are then collected over a time window $T_2^+ = 1000$ to ensure statistical convergence of the mean velocity and the Reynolds stresses.

We note that the pressure gradient is such that the mass flow is the one that led to the desired $Re_{\tau}$ in DNS \cite{LM:2015}. As a consequence, the $Re_{\tau}$ computed from the wall stress in LES (denoted by $Re_{\tau}^{LES}$ hereinafter) may not exactly agree with the target $Re_{\tau}$ (denoted by $Re_{\tau}^{target}$) if the resolution is not sufficiently fine to match the wall stress in DNS. This will be the case at $Re_{\tau}^{target} = 544$ and, for some of the explicit models, also at $182$.

\begin{table}[h]
\centering
\begin{tabular}{ccc}
\hline
 & $Re_{\tau}^{target} = 182$ & $Re_{\tau}^{target} = 544$ \\
\hline
$\Delta x^+$ & $ 15.9 $ & $ 47.5 $ \\
$\Delta y^+_{w}$ & $ 0.29 $ & $ 0.87 $ \\
$\Delta y^+_{avg}$ & $ 3.79 $ & $ 11.3 $ \\
$\Delta z^+$ & $ 9.53 $ & $ 28.5 $ \\
\hline
\end{tabular}
\caption{\label{infoChannelDiscr} Distance between high-order nodes (in wall units) for turbulent channel flow. $\Delta y^+_{avg}$ denotes the average distance along the wall-normal direction. $\Delta y^+_w$ denotes the distance from the wall to the first high-order node along the wall-normal direction.}
\end{table}


\subsection{Numerical results}



Prior to presenting the numerical results, we note that homogeneity in the $x$ and $z$ directions holds pointwise in the exact solution, but only {\it elementwise} in the DG solution. We shall omit this nuance and compute ensemble averages (i.e.\ time averages) as time-, streamwise- and spanwise- averages. This accelerates convergence of the statistics of the flow, and allows comparison with DNS results for which the ensemble-averaged quantities are only a function of the wall-normal coordinate.

\subsubsection{Riemann solver study}

\begin{figure}
\centering
\includegraphics[width=0.49\textwidth]{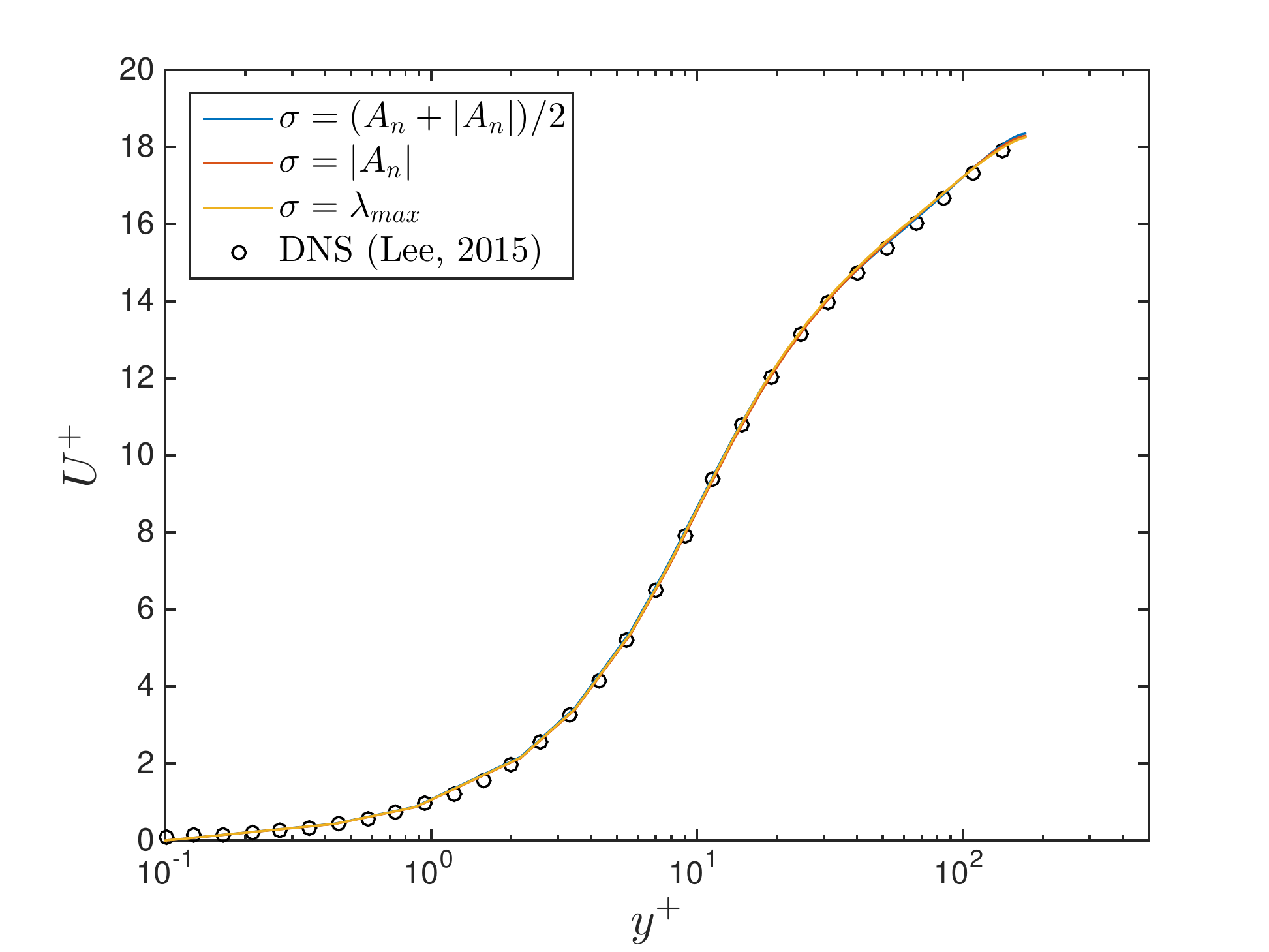}
\hfill \includegraphics[width=0.49\textwidth]{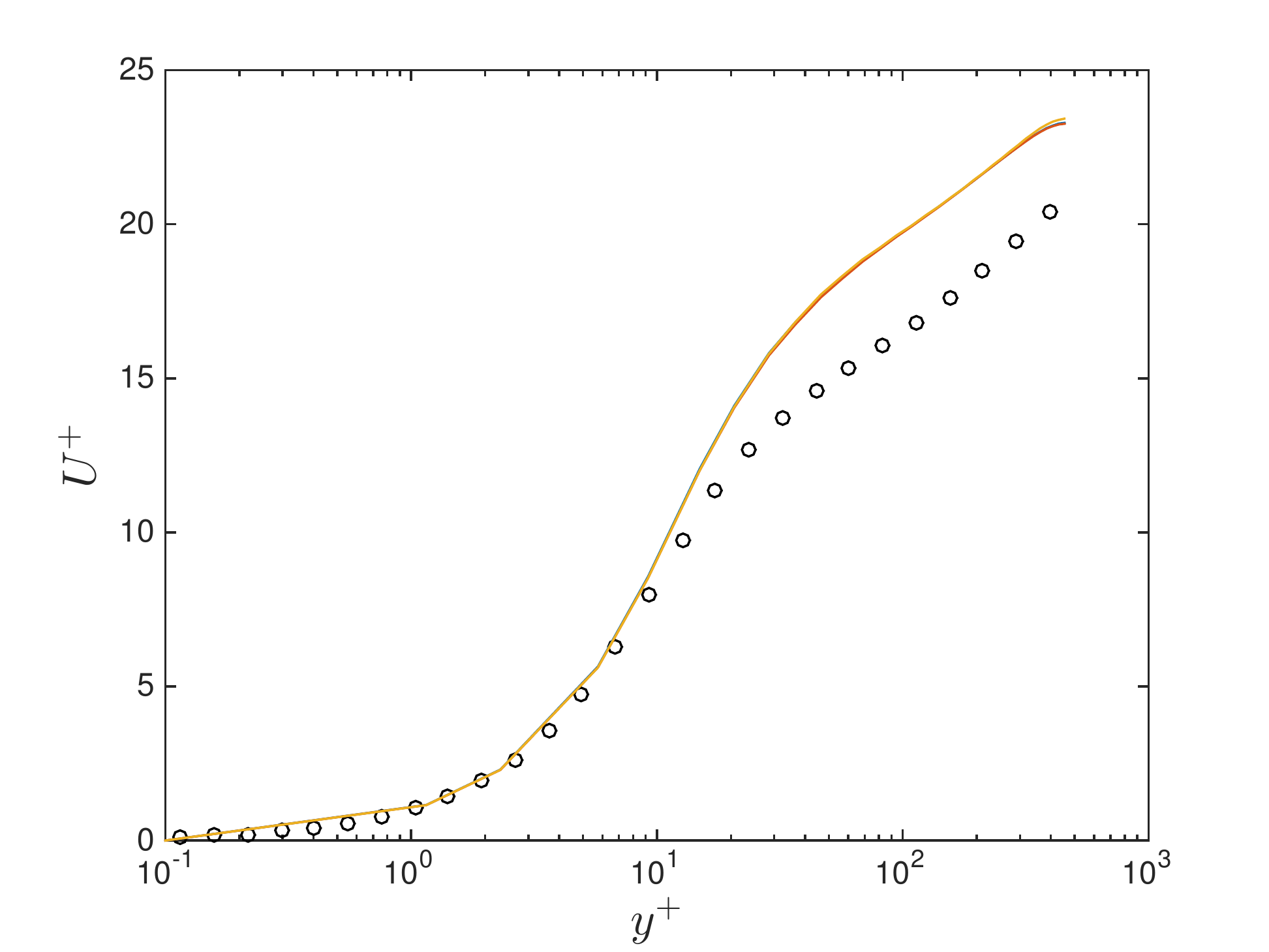}
\caption{\label{meanVel_Riemann} Mean velocity profile in turbulent channel flow at $Re_{\tau}^{target} = 182$ (left) and $544$ (right) for the Riemann solvers considered. The wall friction in the simulations is used to compute the near-wall velocity and length scales for non-dimensionalization.}
\end{figure}

\begin{figure}[h]
\centering
    \addtolength{\leftskip} {-1.5cm}
    \addtolength{\rightskip}{-1.5cm}
\includegraphics[width=1.1\textwidth]{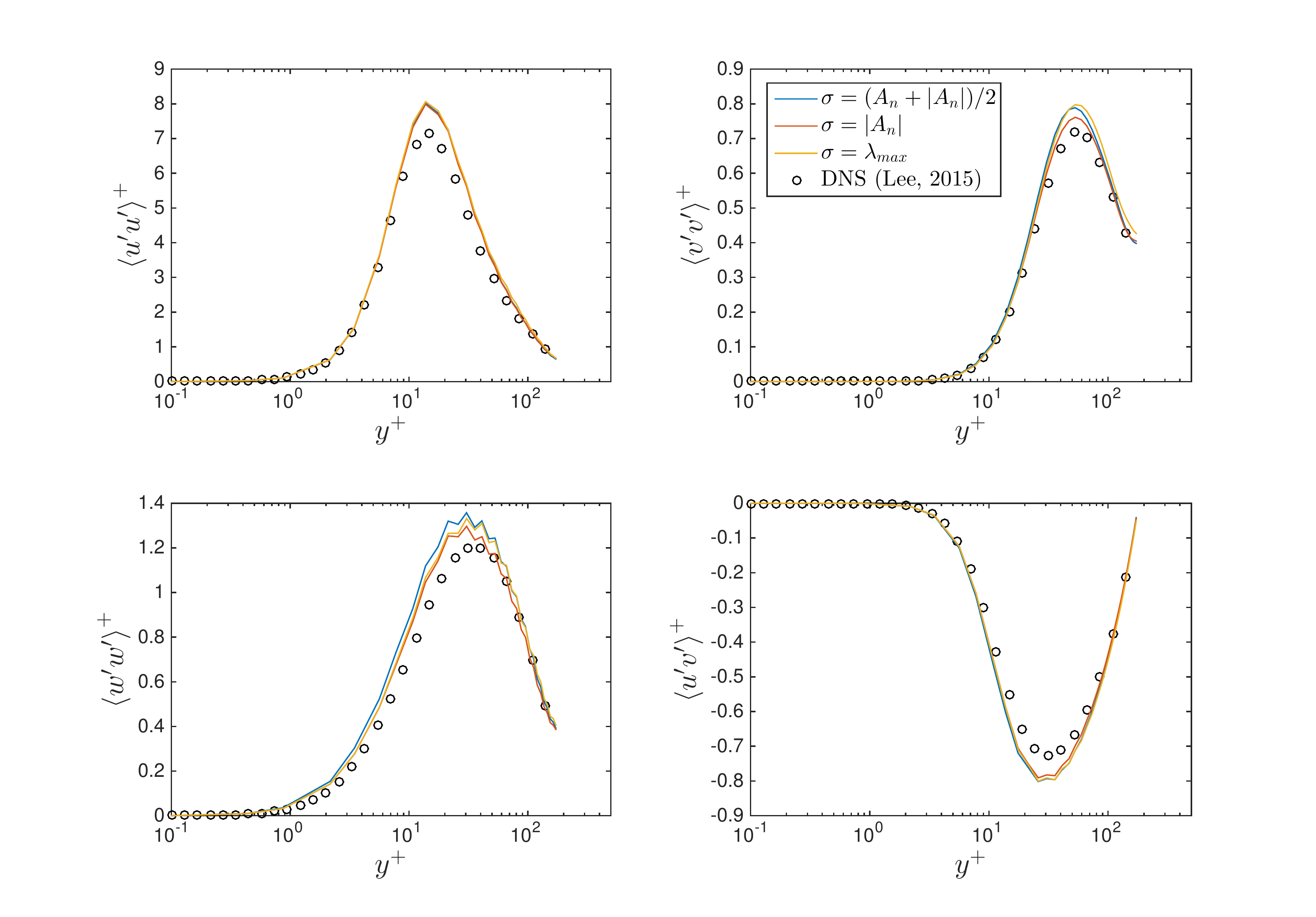}
\caption{\label{ReyStressRe182_Riemann} Reynolds stresses in turbulent channel flow at $Re_{\tau}^{target} = 182$ for the Riemann solvers considered. The wall friction in the simulations is used to compute the near-wall velocity and length scales for non-dimensionalization.}
\end{figure}

Figure \ref{meanVel_Riemann} shows the mean velocity profile at $Re_{\tau}^{target} = 182$ and $544$ for the Riemann solvers considered. The Reynolds stresses at $Re_{\tau}^{target} = 182$ are shown in Figure \ref{ReyStressRe182_Riemann}, and the values of $Re_{\tau}^{LES}$ at both Reynolds numbers are collected in Table \ref{computed_Re_tau}. There are no significant differences between Riemann solvers in terms of wall friction, mean velocity and Reynolds stresses. Further analysis of these results is presented in the comparison between implicit and explicit models below. 




\subsubsection{\label{s:channel_SGS}Subgrid-scale model study}

\begin{figure}[t!]
\centering
\includegraphics[width=0.49\textwidth]{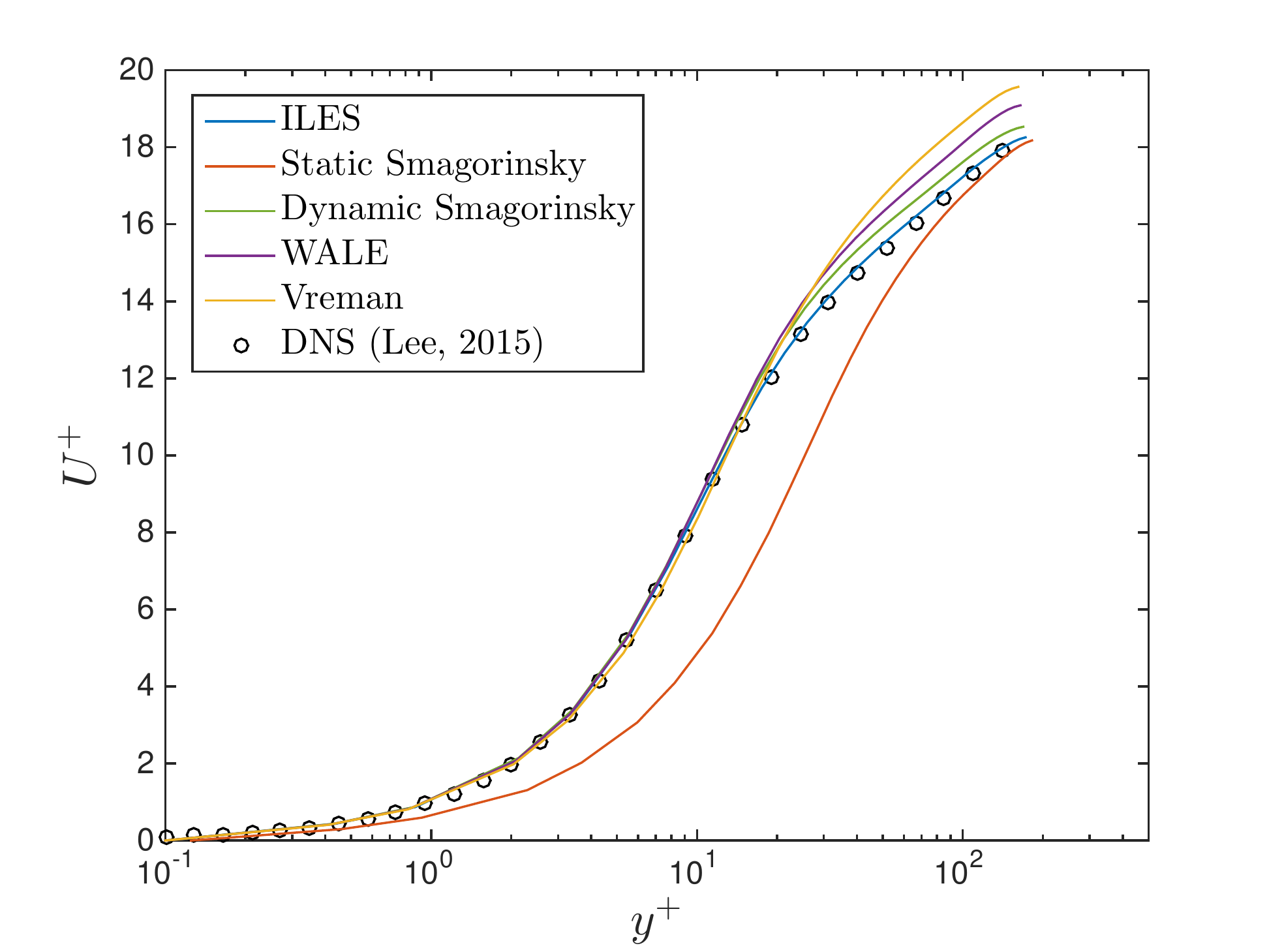}
\hfill \includegraphics[width=0.49\textwidth]{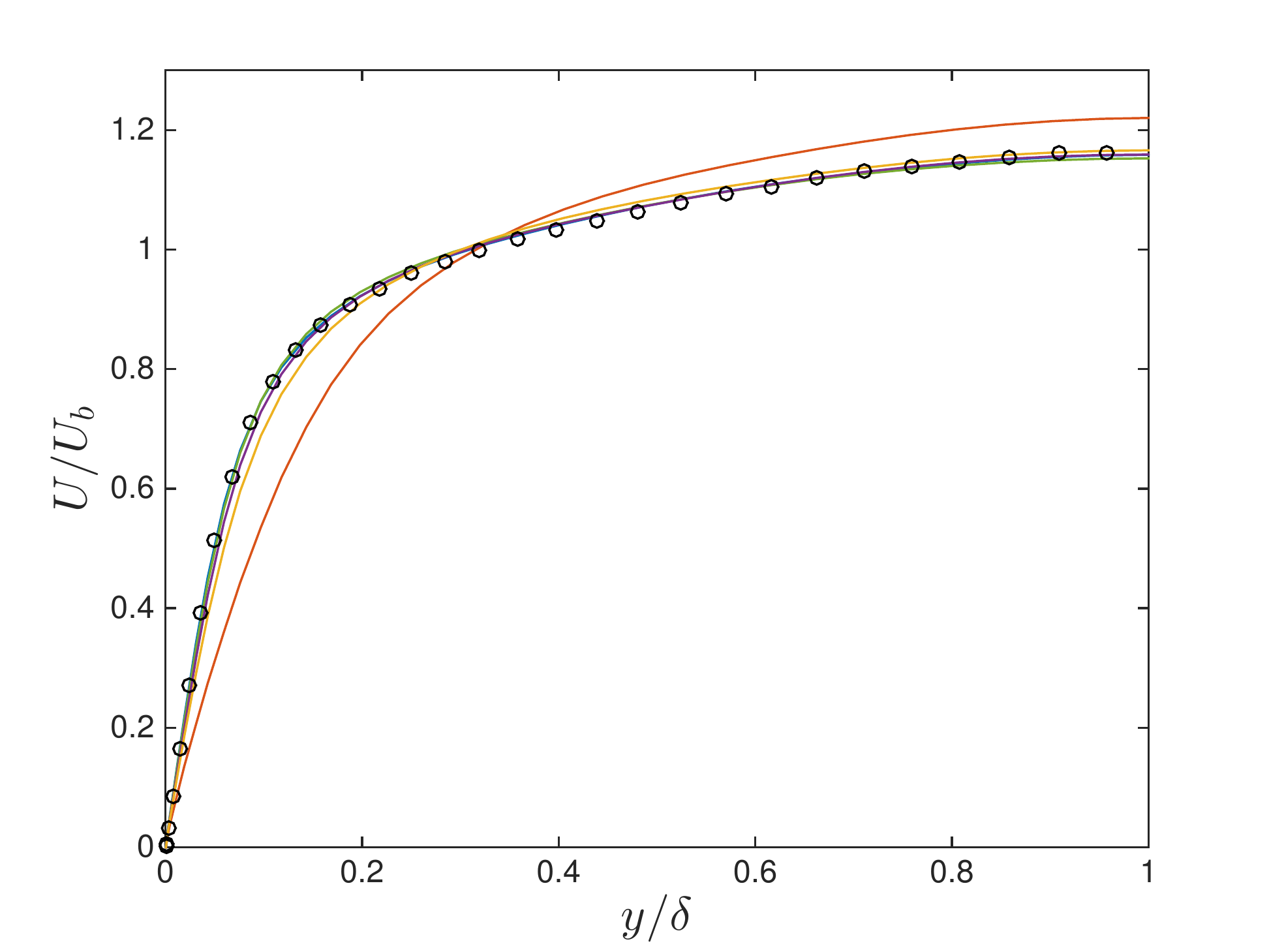}
\caption{\label{f:meanVelRe182_SGS} Mean velocity profile in turbulent channel flow at $Re_{\tau}^{target} = 182$ for the SGS models considered. Inner (based on the wall friction in the simulation) and outer boundary layer scales are used for non-dimensionalization in the left and right figures, respectively.}
\end{figure}

\begin{figure}[t!]
\centering
\includegraphics[width=0.49\textwidth]{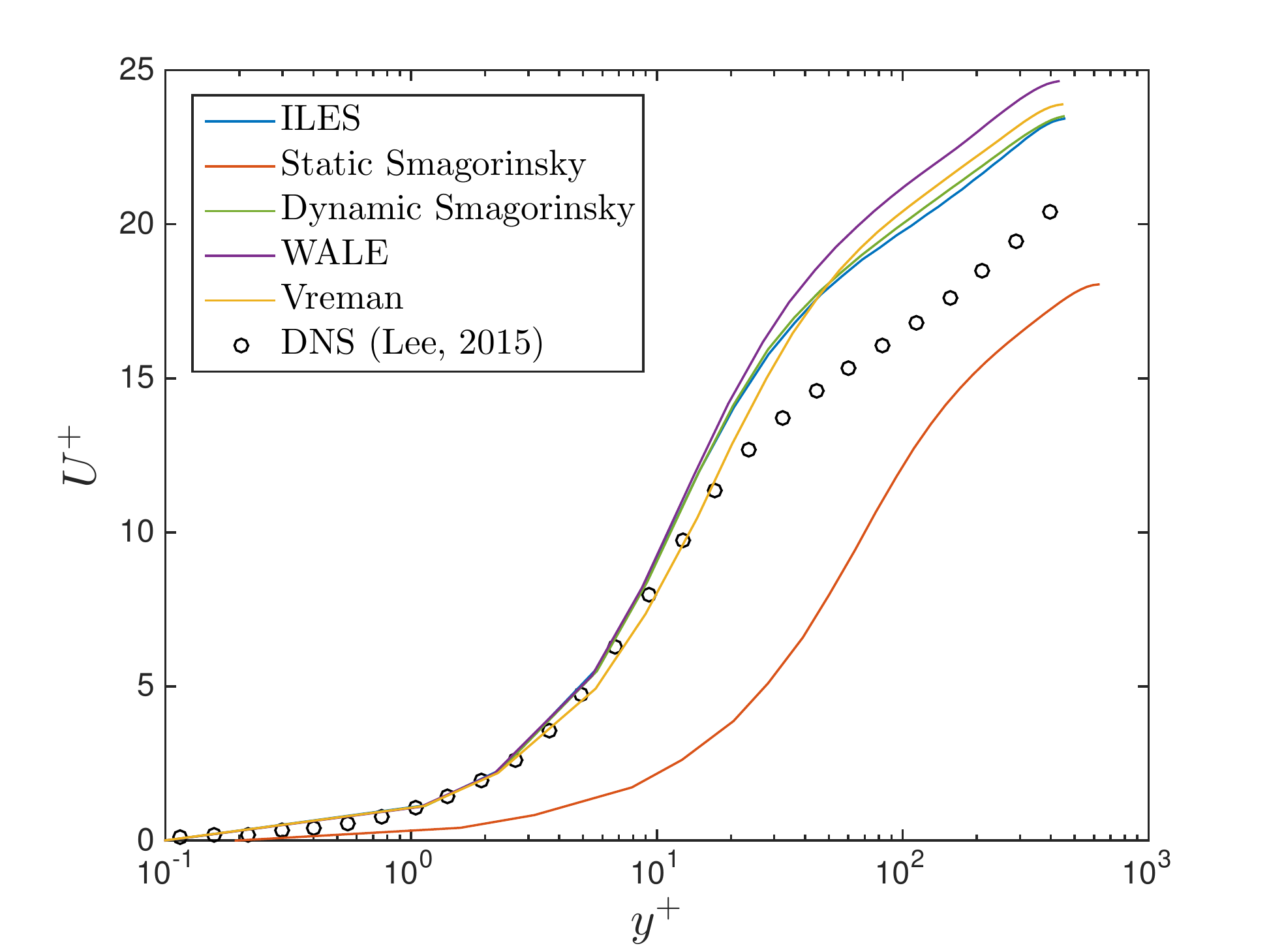}
\hfill \includegraphics[width=0.49\textwidth]{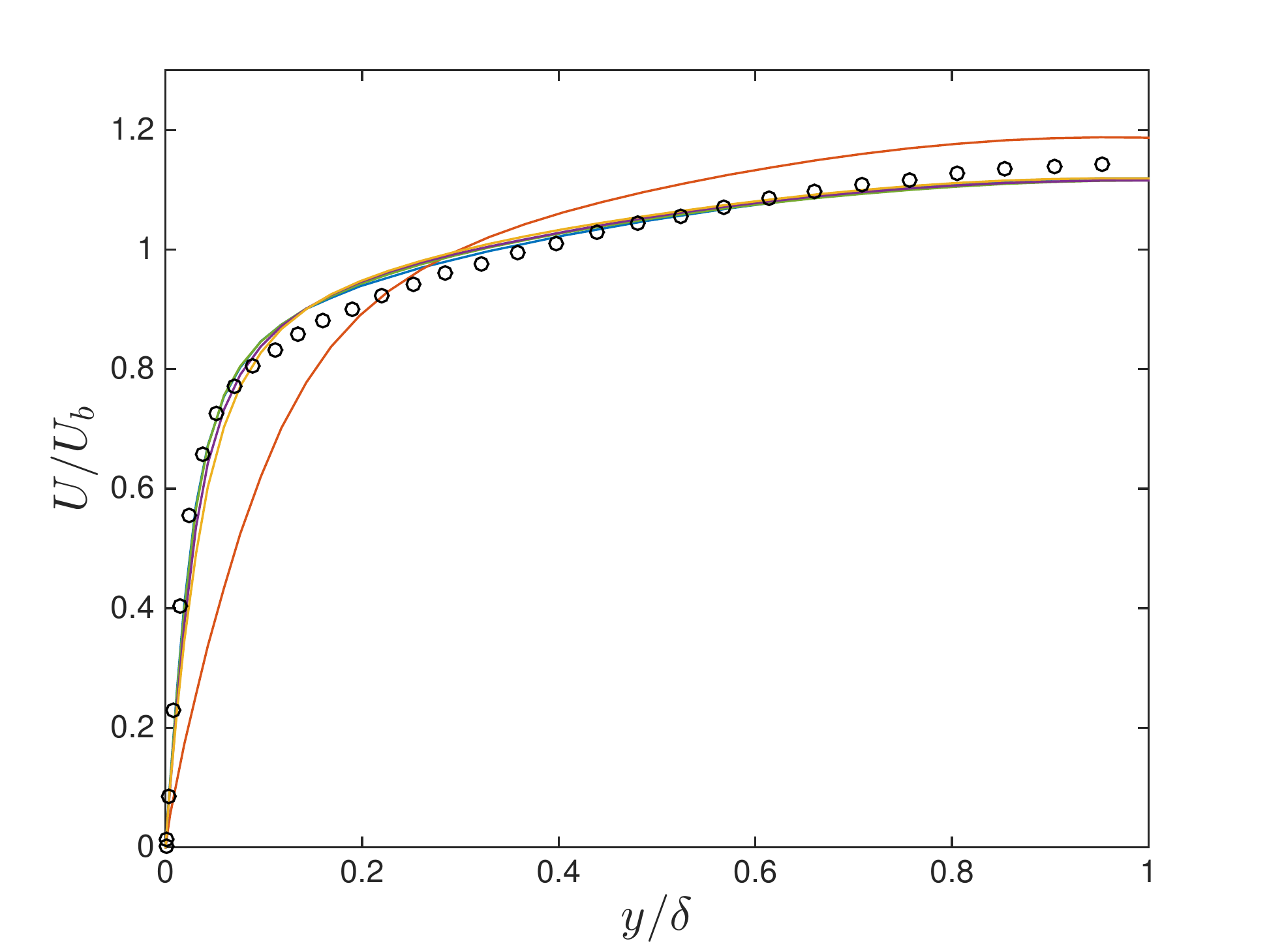}
\caption{\label{f:meanVelRe544_SGS} Mean velocity profile in turbulent channel flow at $Re_{\tau}^{target} = 544$ for the SGS models considered. Inner (based on the wall friction in the simulation) and outer boundary layer scales are used for non-dimensionalization in the left and right figures, respectively.}
\end{figure}

Figures \ref{f:meanVelRe182_SGS} and \ref{f:meanVelRe544_SGS} show the mean velocity profile at $Re_{\tau}^{target} = 182$ and $544$, respectively, with ILES, the static Smagorinsky, dynamic Smagorinsky, WALE and Vreman models. Two different types of non-dimensionalization are used on the left and right images. On the left images, the inner boundary layer scales, namely the near-wall velocity and length scales, are used for non-dimensionalization. These scales are computed from the wall friction in the simulations, as opposed to the target wall friction. 
This is the proper choice to investigate if the viscous, buffer and log sublayers of the boundary layer are accurately resolved, as we will discuss below. 
On the right images, the outer boundary layer scales, namely $U_b$ and $\delta$, are used for non-dimensionalization so that all simulations are non-dimensionalized with respect to the same reference values. This way, the scaling factor between dimensional and non-dimensional data is the same in all cases, and the dimensional velocity profiles can be directly compared. Both choices of non-dimensionalization complement each other and help understand the performance of the models and the reasons for the mismatch (if any) between LES and DNS.

\begin{figure}[t!]
\centering
    \addtolength{\leftskip} {-1.5cm}
    \addtolength{\rightskip}{-1.5cm}
\includegraphics[width=1.1\textwidth]{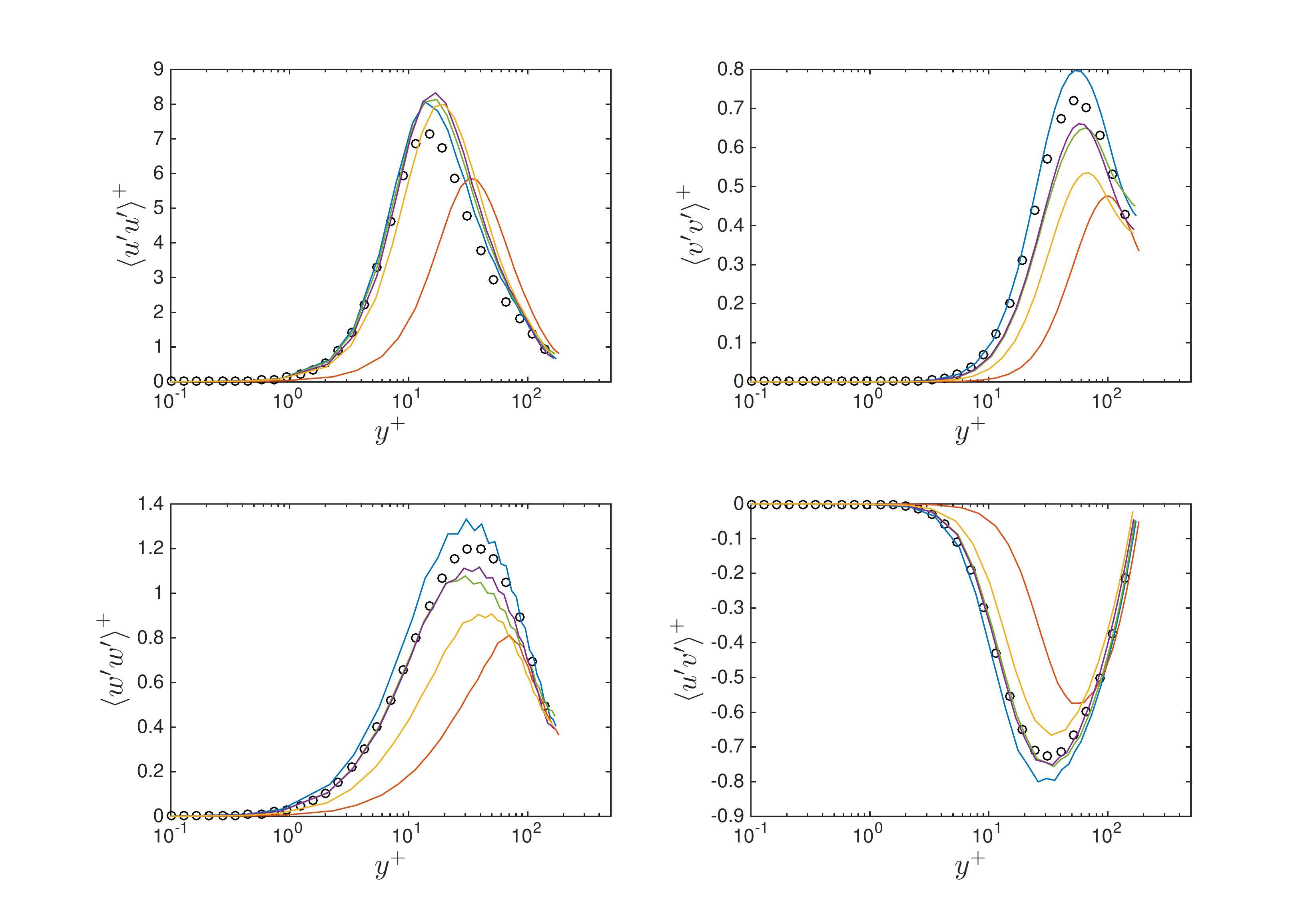}
\caption{\label{ReyStressRe182_SGS} Reynolds stresses in turbulent channel flow at $Re_{\tau}^{target} = 182$ for the SGS models considered. The wall friction in the simulations is used to compute the near-wall velocity and length scales for non-dimensionalization. The color legend is shown in Figures \ref{f:meanVelRe182_SGS} and \ref{f:meanVelRe544_SGS}.}
\end{figure}

\begin{figure}
\centering
\includegraphics[width=0.49\textwidth]{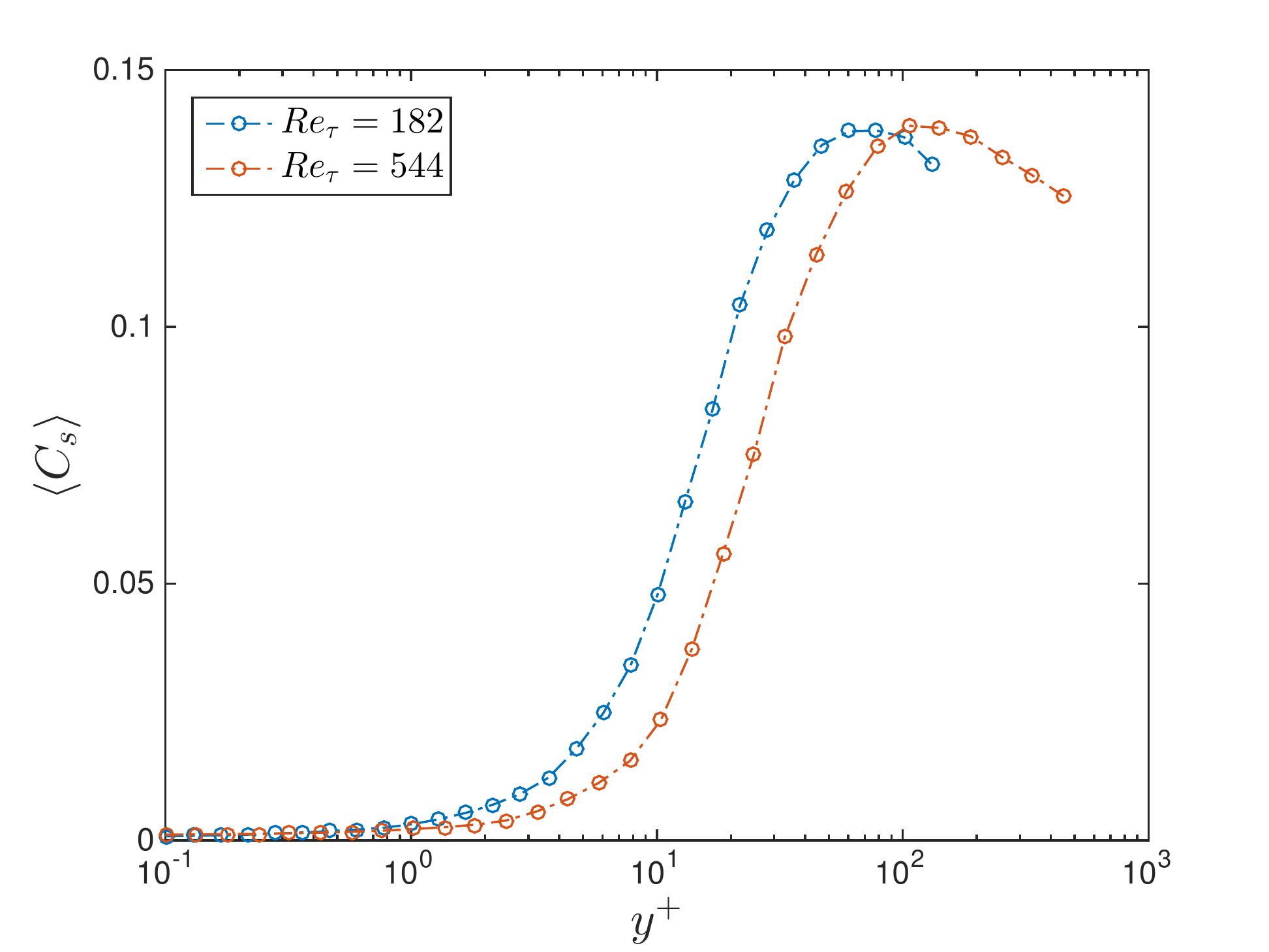}
\caption{\label{f:Cs_dynSmag_channel} Dynamic Smagorinsky constant in turbulent channel flow at $Re_{\tau}^{target} = 182$ and $544$. Note $C_s = 0.16$ in static Smagorinsky \cite{Gatski:09}.}
\end{figure}

Figure \ref{ReyStressRe182_SGS} shows the Reynolds stresses at $Re_{\tau}^{target} = 182$, Figure \ref{f:Cs_dynSmag_channel} the dynamic Smagorinsky constant at both Reynolds numbers, and Table \ref{computed_Re_tau} the values of $Re_{\tau}^{LES}$ at both Reynolds numbers. Prior to discussing these results, we introduce the two following quantities
\begin{subequations}
\label{e:mue}
\begin{alignat}{1}
\mu_e^r & := - \mu \, \langle u' v' \rangle^+ \bigg( \frac{d U^+}{d y^+} \bigg) ^{-1} , \\
\mu_e^* & := \mu \, \bigg[ \bigg( 1 - \frac{y^+}{Re_{\tau}} \bigg) \bigg( \frac{d U^+}{d y^+} \bigg) ^{-1} - 1 \bigg] . 
\end{alignat}
\end{subequations}
From these definitions, it follows that $\mu_e^r \, dU / dy = - \rho_0 \langle u' v' \rangle$ and $(\mu+\mu_e^*) \, dU / dy = \tau_w (1 - y / \delta)$. Note that the mean shear stress varies linearly across the channel due to the uniform pressure gradient and $x$-momentum conservation, 
and therefore $\tau_w (1-y/\delta) = \langle \tau_{xy} \rangle$. Note also that $- \rho_0 \langle u' v' \rangle \approx - \langle \rho u'' v'' \rangle$, where $u''$ and $v''$ are the Favre fluctuating velocities, and thus this term corresponds to the net resolved $x$-momentum turbulent transport, per unit area, across $y$-planes. From these considerations, $\mu_e^r$ can be interpreted as an eddy viscosity due to the resolved turbulent motion, and will be referred to as the {\it resolved eddy viscosity}; whereas $\mu_e^*$ can be regarded as a {\it total eddy viscosity} or an {\it effective eddy viscosity} in the simulation, and 
accounts for the turbulent transport due to the explicit model, the implicit model and the resolved turbulence. 
Note that, for the exact solution, it holds that $\mu_e^* = \mu_e^r$. Also, a {\it modeled eddy viscosity}, accounting for the implicit and explicit models, is not straightforward to define in the context of DG methods and in fact it cannot be computed as $\mu_e^* - \mu_e^r$.

\begin{figure}
\centering
\hfill \includegraphics[width=0.49\textwidth]{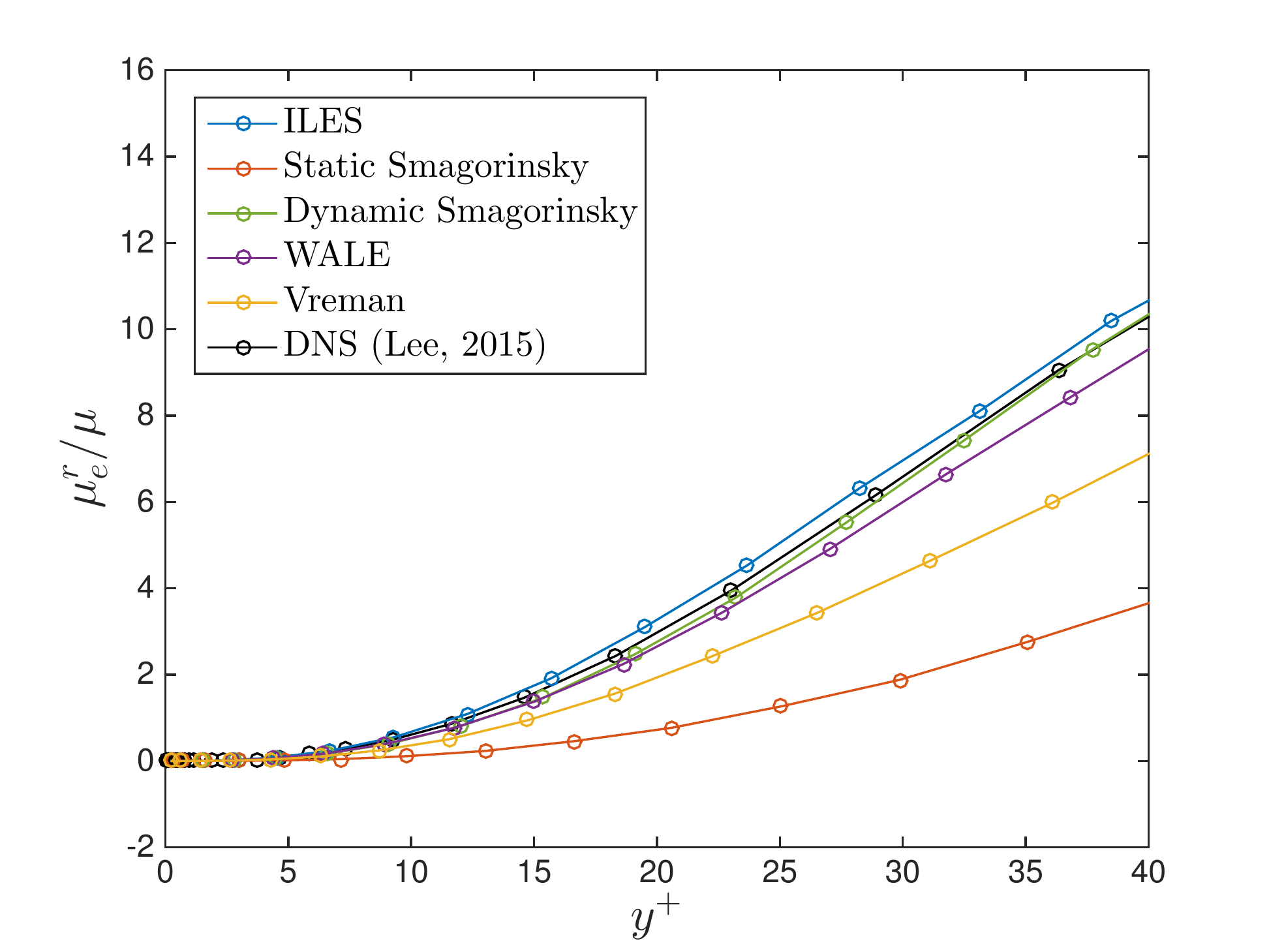}
\hfill \includegraphics[width=0.49\textwidth]{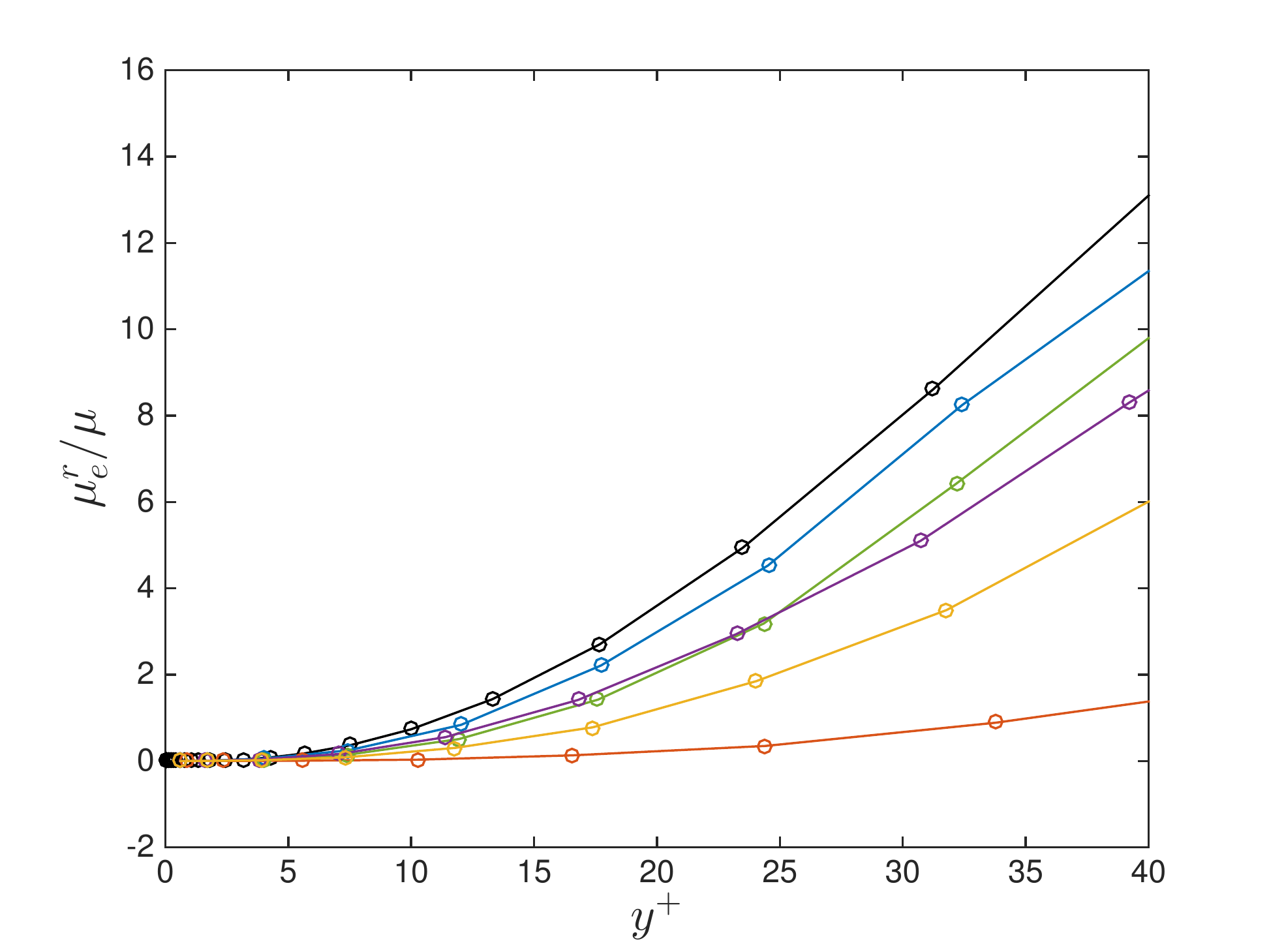}
\includegraphics[width=0.49\textwidth]{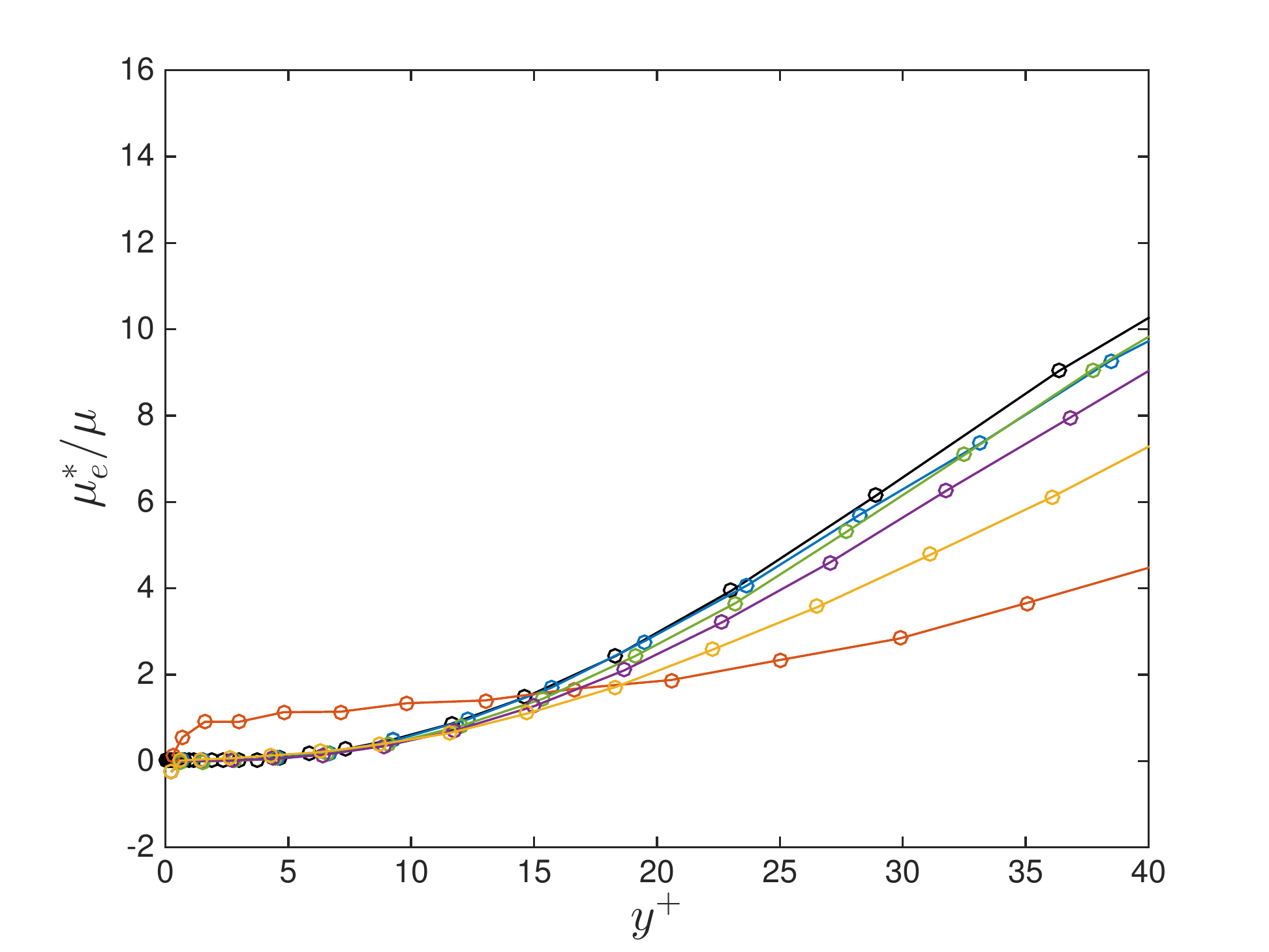}
\hfill \includegraphics[width=0.49\textwidth]{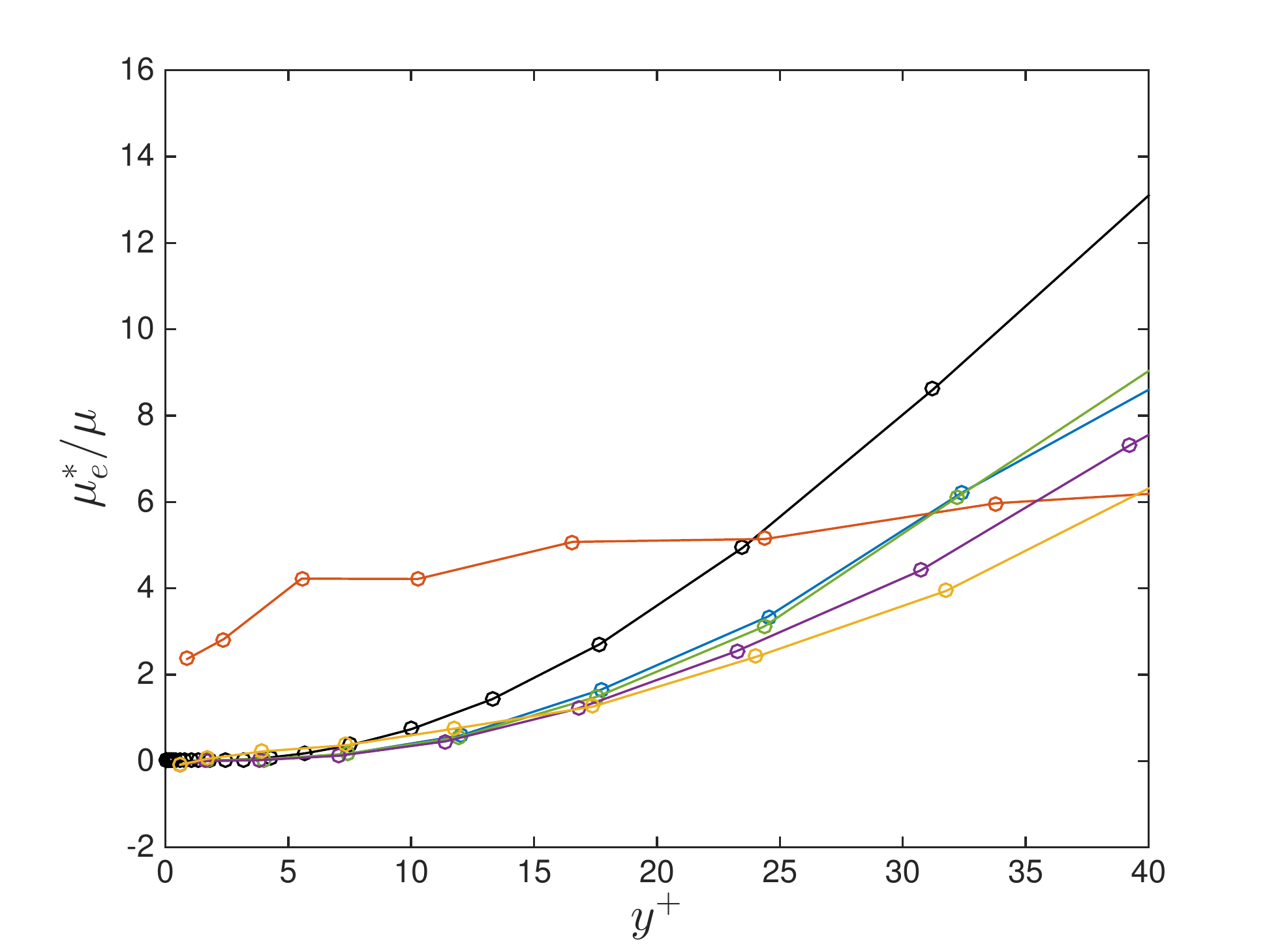}
\caption{\label{f:nuE_channel} Resolved eddy viscosity (top) and total eddy viscosity (bottom) in turbulent channel flow at $Re_{\tau}^{target} = 182$ (left) and $544$ (right) for the SGS models considered. The definition of these eddy viscosities is given in Eq. \eqref{e:mue}. The wall friction in the simulations is used to compute the near-wall velocity and length scales for non-dimensionalization.}
\end{figure}

The resolved eddy viscosity in the viscous and buffer layers for the SGS models considered is shown in the top of Figure \ref{f:nuE_channel}. The total eddy viscosity is shown in the bottom of the figure. Except for static Smagorinsky, $\mu_e^*$ vanishes as expected in the viscous layer, and grows in the buffer layer as the turbulent transport increases and dominates the molecular transport. We emphasize that these eddy viscosities inform of the turbulent transport, and not of the dissipation of kinetic energy which was the focus in the Taylor-Green vortex. While eddy viscosity SGS models account for the subgrid scale contribution to these two terms through an (the same) explicit eddy viscosity, these terms are of a fundamentally different nature, as apparent from the filtered momentum and kinetic energy equations \cite{Garnier:2009}.

All the ingredients are now in place to discuss the numerical results. We focus first on the viscous sublayer, then on the buffer layer, and finally on the log layer.

\begin{table}[t!]
\centering
\begin{tabular}{ccc}
\hline
 & $Re_{\tau}^{target} = 182$ & $Re_{\tau}^{target} = 544$ \\
\hline
\multicolumn{3}{c}{Study of the Riemann solver} \\
\hline
$(A_n + |A_n|)/2$ & $ 180.8 $ & $ 478.7 $ \\
$|A_n|$ & $ 181.0 $ & $ 479.1 $ \\
$\lambda_{max}$ & $ 181.1 $ & $ 477.7 $ \\
\hline
\multicolumn{3}{c}{Study of the SGS model} \\
\hline
ILES & $ 180.8 $ & $ 478.7 $ \\
Static Smagorinsky & $ 191.5 $ & $ 658.1 $ \\
Dynamic Smagorinsky & $177.6$ & $474.4$ \\
WALE & $ 173.3 $ & $ 452.6 $ \\
Vreman & $ 170.1 $ & $ 468.4 $ \\
\hline
\end{tabular}
\caption{\label{computed_Re_tau} Values of $Re_{\tau}^{LES}$ computed from the wall friction in the LES simulations.}
\end{table}



 
\noindent {\bf Viscous sublayer ($y^+ \lesssim 10$):} This layer is characterized by $\langle \tau_{xy} \rangle \approx \tau_w$ and the fact that molecular transport dominates turbulent transport; which yields the well-known result $U^+ \approx y^+$. Since this velocity law is independent of the Reynolds number, all simulations should ideally match the DNS data on the left images of Figures \ref{f:meanVelRe182_SGS}$-$\ref{f:meanVelRe544_SGS}, despite the slightly different Reynolds number in LES. From these two figures, the viscous sublayer is accurately resolved at both Reynolds numbers by implicit LES, the dynamic Smagorinsky, the WALE and the Vreman models. Provided that there is enough resolution to capture the mean flow and that no eddy viscosity is added by the SGS model, this is expected due to the lack of subgrid scales in this laminar sublayer. The eddy viscosity, however, does not vanish near the wall with static Smagorinsky (see $\mu_e^*$ in Figure \ref{f:nuE_channel}) and this leads to inaccurate wall friction and mean velocity in this layer. 
While no significant differences were observed between static Smagorinsky and the other static models, namely WALE and Vreman, for the wall-free flow conditions in the Taylor-Green vortex, this is not the case for this wall-bounded flow and, as expected, points to the inability of the static Smagorinsky model to vanish near walls as one the main limitations of the model.

\noindent {\bf Buffer layer ($10 \lesssim y^+ \lesssim 40$):} 
The mean velocity $U^+ = U^+(y^+)$ in the buffer layer is also approximately independent of the Reynolds number \cite{Bernardini:2014,Alamo:2004,Hoyas:2006,LM:2015,Lozano:2014,Moser:1999}, and therefore LES should ideally match DNS on the left images of Figures \ref{f:meanVelRe182_SGS}$-$\ref{f:meanVelRe544_SGS}. 
At $Re_{\tau}^{target} = 182$, the buffer layer is only accurately resolved with ILES. At $Re_{\tau}^{target} = 544$, the buffer layer is not accurately resolved in any of the LES simulations. 
From the top images in Figure \ref{f:nuE_channel}, the use of an explicit model reduces the resolved turbulent transport. As discussed before and justified by linear analysis techniques \cite{Alhawwary:2018,Moura:15a,Fernandez:nonModal:2018}, 
the explicit models damp the large scales more than the implicit model does. 
Since the large turbulent scales are responsible for most of the turbulent transport, this justifies the smaller $\mu_e^r$. 
From the bottom images in Figure \ref{f:nuE_channel}, the total eddy viscosity $\mu_e^*$ is also smaller with an explicit model, and so is the total turbulent transport.

\noindent {\bf Log layer ($y^+ \gtrsim 40$):} This layer is characterized by $\langle \tau_{xy} \rangle \approx \tau_w$ and the fact that turbulent transport dominates molecular transport. This, combined with the mixing-length hypothesis $\mathcal{\ell} = \kappa \ d_w$, where $\mathcal{\ell}$ is the mixing length, $\kappa$ is the von K\'arm\'an constant and $d_w$ is the distance to the wall, leads to $U^+ \approx \kappa^{-1} \log{y^+} + B$, where $\kappa , B > 0$ are positive constants that, for channel flows, are approximately independent of the Reynolds number. DNS \cite{Bernardini:2014,Alamo:2004,Hoyas:2006,LM:2015,Lozano:2014,Moser:1999} and experimental \cite{Monty:2009,Schultz:2013} data have confirmed the existence of this log layer, and in particular 
$\kappa \approx 0.384$ and $B \approx 4.27$ \cite{LM:2015}. The left images in Figures \ref{f:meanVelRe182_SGS} and \ref{f:meanVelRe544_SGS} show that all simulations succeed to predict this logarithmic dependence with the correct mixing length $\mathcal{\ell} = \kappa_{LES} \ d_w , \ \kappa_{LES} \approx 0.38$. The matching velocity between the buffer and log layers, however, does not agree with DNS, and thus $B_{LES} \neq 4.27$, if the buffer layer is not accurately resolved. 
This leads to an incorrect outer velocity seen by the viscous sublayer; which is responsible for the misprediction of the wall friction and thus $Re_{\tau}^{LES}$ in Table \ref{computed_Re_tau}.


To conclude, we note that the Reynolds stresses 
have a stronger dependence on the Reynolds number \cite{LM:2015}. Since the simulations with different SGS models predict slightly different wall frictions, and thus 
solve slightly different Reynolds numbers, it is challenging to infer additional conclusions, beyond those already discussed, from the Reynolds stresses in Figures \ref{ReyStressRe182_Riemann} and \ref{ReyStressRe182_SGS}.

\section{\label{s:conclusions}Conclusions}

We investigated the ability of discontinuous Galerkin methods to predict under-resolved turbulent flows. The Taylor-Green vortex and the turbulent channel flow at various Reynolds numbers were considered to this end. Numerical results showed that DG methods without an explicit subgrid-scale model (implicitly) introduce numerical dissipation in under-resolved turbulence simulations. This implicit {\it subgrid-scale model} is due to the inter-element jumps and the Riemann solver, and behaves like a dynamic model in the sense that it vanishes for laminar flows that do not contain subgrid scales; which is a critical feature to accurately simulate transitional flows. 
In addition, for the moderate-Reynolds-number turbulence 
problems considered, the implicit model provided a more accurate representation of the actual subgrid scales than 
state-of-the-art explicit eddy viscosity models.
Theoretical results for DG methods were used to justify these numerical observations.

Some premises that are widely accepted in the LES community do no longer hold in the context of high-order DG methods. First, the built-in subgrid-scale model in the DG scheme is partially inhibited when using an explicit model, and the total amount
of dissipation 
does not necessarily increase with an explicit model. Second, since eddy viscosity models dissipate kinetic energy at larger scales than the implicit model, they reduce more significantly the energy content of scales that are larger than the grid Nyquist wavenumber; which may have important consequences in practice. 
In particular, explicit eddy viscosity models may not allow taking advantage of the low numerical dissipation at large scales of high-order DG methods \cite{Fernandez:nonModal:2018,Moura:15a}; which is critical for transition prediction and moderate-Reynolds-number turbulence. From these two observations, a change in the current best practices for subgrid-scale modeling may be required with DG methods. 
This is not completely surprising considering that the state-of-the-art subgrid-scale models, based on the Boussinesq eddy viscosity assumption and an augmented viscous operator, have been developed and successfully applied with discretization schemes whose built-in stabilization (if any) and dissipation characteristics 
are different from those in discontinuous Galerkin methods. 

To conclude, we briefly discuss how our results are expected to extend to higher accuracy orders. For accuracy orders up to about seven, similar results to those in this paper are expected. For accuracy orders beyond about eight, the numerical dissipation in the DG scheme becomes very small also at high wavenumbers \cite{Fernandez:nonModal:2018}, and some form of regularization/model may be required to enhance stability and accuracy. It will be critical that the explicit regularization/model used with high-order DG schemes localizes dissipation at the desired wavenumbers and is consistent with the expected SGS dissipation spectrum \cite{Cerutti:2000,Kraichnan:1976,Langford:1999}; which could be achieved with Spectral Vanishing Vicosity \cite{Tadmor:1989,Karamanos:2000,Kirby:2006,Moura:16b}, Variational Multiscale \cite{Collis:2002,Hughes:1998,Hughes:2000} and Mori-Zwanzig \cite{Parish:2017,Parish:2017b,Parish:MZVMS:2018} type approaches. 

\section*{Acknowledgments}

The authors acknowledge Pratt \& Whitney, the Air Force Office of Scientific Research (FA9550-16-1-0214) and the National Aeronautics and Space Administration (NASA NNX16AP15A) for supporting this work. The first author also acknowledges the financial support from the Zakhartchenko and ``la Caixa'' Fellowships.

\appendix

\section{\label{s:HDGdiscretizationApp}The hybridized discontinuous Galerkin methods for the compressible Euler and Navier-Stokes equations}

\subsection*{\label{s:governingEq}Governing equations}

Let $t_f > 0$ be a final time and let $\Omega \subset \mathbb{R}^d, \, 1 \leq d \leq 3$ be an open, connected and bounded physical domain with Lipschitz boundary $\partial \Omega$. The unsteady, compressible Navier-Stokes equations in conservation form read as
\begin{subequations}
\label{e:NS}
\begin{alignat}{2}
\label{e:ns0}
\displaystyle \bm{q} - \nabla \bm{u} = 0 & , \qquad \mbox{in } \Omega \times [0, t_f) ,   \\
\label{e:ns1}
\displaystyle \frac{\partial  \bm{u}}{\partial t} +  \nabla  \cdot  \bm{F}(\bm{u}) +  \nabla  \cdot  \bm{G}(\bm{u} , \bm{q}) = 0 & , \qquad \mbox{in } \Omega \times [0, t_f) ,   \\
\label{e:ns2}
\bm{B}(\bm{u} , \bm{q}) = 0 & , \qquad \mbox{on } \partial \Omega \times [0,t_f) , \\
\label{e:ns3}
\bm{u} - \bm{u}_0 = 0 & , \qquad \mbox{on } \Omega \times \{ 0 \} . 
\end{alignat}
\end{subequations}
Here, $\bm{u} = (\rho, \rho v_{j}, \rho E) \in \mathbb{R}^m, \, \ j=1,...,d$ is the $m$-dimensional ($m = d + 2$) vector of conservation variables, $\bm{u}_0$ is an initial condition, $\bm{B}(\bm{u},\bm{q})$ is a boundary operator, and $\bm{F}(\bm{u})$ and $\bm{G}(\bm{u} , \bm{q})$ are the inviscid and viscous fluxes of dimensions $m \times d$, given by
\begin{equation}
\label{flux}
\bm{F}(\bm{u}) = \left( \begin{array}{c}
\rho v_j \\
\rho v_i v_j + \delta_{ij} p \\
 v_j (\rho E + p)
\end{array}
\right) , \qquad
\bm{G}(\bm{u},\bm{q}) = - \left( \begin{array}{c}
0 \\
\tau_{ij}  \\
v_i \tau_{ij} - f_j
\end{array}
\right) , \qquad i , j = 1 , \dots , d , 
\end{equation}
where $p$ is the thermodynamic pressure, $\tau_{ij}$ the viscous stress tensor, $f_j$ the heat flux, and $\delta_{ij}$ the Kronecker delta. For a calorically perfect gas in thermodynamic equilibrium, $p = (\gamma - 1) \, \big( \rho E - \rho \, \norm{\bm{v}}^2 / 2 \big)$, where $\gamma = c_p / c_v > 1$ is the ratio of specific heats and in particular $\gamma \approx 1.4$ for air. $c_p$ and $c_v$ are the specific heats at constant pressure and volume, respectively. For a Newtonian fluid with the Fourier's law of heat conduction, the viscous stress tensor and heat flux are given by
\begin{equation}
\label{e:closuresNS}
\tau_{ij} = \mu \, \bigg( \frac{\partial v_i}{\partial x_j}+\frac{\partial v_j}{\partial x_i} - \frac{2}{3}\frac{\partial v_k}{\partial x_k}\delta_{ij} \bigg) + \beta \, \frac{\partial v_k}{\partial x_k}\delta_{ij} , \qquad \qquad f_j = - \, \kappa \, \frac{\partial T}{\partial x_j} , 
\end{equation}
where $T$ denotes temperature, $\mu$ the dynamic (shear) viscosity, $\beta$ the bulk viscosity, $\kappa = c_p \, \mu / Pr$ the thermal conductivity, and $Pr$ the Prandtl number. In particular, $Pr \approx 0.71$ for air, and additionally $\beta = 0$ under the Stokes' hypothesis. The unsteady, compressible Euler equations are obtained by dropping the viscous flux in Eq. \eqref{e:ns1}.


\subsection*{\label{s:FEmesh}Finite element mesh}

We denote by $\mathcal{T}_h$ a collection of stationary, non-singular, conforming, $p$-th degree curved elements $K$ that partition\footnote{Strictly speaking, the finite element mesh can only partition the problem domain if $\partial \Omega$ is piecewise $p$-th degree polynomial. For simplicity of exposition, and without loss of generality, we assume that the elements in $\mathcal{T}_h$ actually partition $\Omega$. In addition, the term {\it partition} actually refers to {\it Lebesgue mod 0 partition}.} $\Omega$, and set $\partial \mathcal{T}_h := \{ \partial K : K \in \mathcal{T}_h \} $ to be the collection of the boundaries of the elements in $\mathcal{T}_h$. For an element $K$ of the collection $\mathcal{T}_h$, $F= \partial K \cap \partial \Omega$ is a boundary face if its $d-1$ Lebesgue measure is nonzero. For two elements $K^+$ and $K^-$ of $\mathcal{T}_h$, $F=\partial K^{+} \cap \partial K^{-}$ is the interior face between $K^+$ and $K^-$ if its $d-1$ Lebesgue measure is nonzero. We denote by $\mathcal{E}_h^I$ and $\mathcal{E}_h^B$ the set of interior and boundary faces, respectively, and we define $\mathcal{E}_h := \mathcal{E}_h^I \cup \mathcal{E}_h^B$ as the union of interior and boundary faces. Note that, by definition, $\partial \mathcal{T}_h$ and $\mathcal{E}_h$ are different. More precisely, an interior face is counted twice in $\partial \mathcal{T}_h$ but only once in $\mathcal{E}_h$, whereas a boundary face is counted once both in $\partial \mathcal{T}_h$ and $\mathcal{E}_h$.

\subsection*{\label{s:FEspaces}Finite element spaces}

Let $\mathcal{P}_{k}(D)$ denote the space of polynomials of degree at most $k$ on a domain $D \subset \mathbb{R}^n$, let $L^2(D)$ be the space of Lebesgue square-integrable functions on $D$, and $\mathcal{C}^0(D)$ the space of continuous functions on $D$. Also, let $\bm{\psi}^p_K$ denote the $p$-th degree parametric mapping from the reference element $K_{ref}$ to an element $K \in \mathcal{T}_h$ in the physical domain, and $\bm{\phi}^p_F$ be the $p$-th degree parametric mapping from the reference face $F_{ref}$ to a face $F \in \mathcal{E}_h$ in the physical domain. We then introduce the following discontinuous finite element spaces in $\mathcal{T}_h$
\begin{subequations}
\begin{alignat}{2}
& \bm{\mathcal{Q}}_{h}^k &&= \big\{\bm{r} \in [L^2(\mathcal{T}_h)]^{m \times d} \ : \ (\bm{r} \circ \bm{\psi}_K^p )  |_K \in [\mathcal{P}_k(K_{ref})]^{m \times d} \ \ \forall K \in \mathcal{T}_h \big\} , \\
& \bm{\mathcal{V}}_{h}^k &&= \big\{\bm{w} \in [L^2(\mathcal{T}_h)]^m \ : \ (\bm{w} \circ \bm{\psi}_K^p )|_K \in [\mathcal{P}_k(K_{ref})]^m \ \ \forall K \in \mathcal{T}_h \big\} , 
\end{alignat}
\end{subequations}
and the following finite element space on the mesh skeleton $\mathcal{E}_{h}$
\begin{equation}
\bm{\mathcal{M}}_{h}^k  = \big\{ \bm{\mu} \in [L^2(\mathcal{E}_h)]^m \ \\
\ : \ (\bm{\mu} \ \circ \ \bm{\phi}^p_F)|_F \in [\mathcal{P}^k(F_{ref})]^m \, \ \forall F \in \mathcal{E}_h , \ \textnormal{and} \ \bm{\mu}|_{\mathcal{E}^{\rm E}_h} \in [C^0(\mathcal{E}^{\rm E}_h)]^m \big\} , 
\end{equation}
where $\mathcal{E}^{\rm E}_h$ is a subset of $\mathcal{E}_h$. Different choices of $\mathcal{E}^{\rm E}_h$ lead to different discretization methods within the hybridized DG family that have different properties in terms of accuracy, stability, and number of globally coupled unknowns \cite{Fernandez:17a,Nguyen:15}. In particular, the HDG, EDG and IEDG methods are obtained by setting $\mathcal{E}^{\rm E}_h = \emptyset$, $\mathcal{E}^{\rm E}_h = \mathcal{E}_h$ and $\mathcal{E}^{\rm E}_h = \mathcal{E}_h^I$, respectively.

We finally define several inner products associated with these finite element spaces. In particular, given $\bm{w}, \bm{v} \in \bm{\mathcal{V}}_{h}^k$, $\bm{W}, \bm{V} \in \bm{\mathcal{Q}}_{h}^k$ and $\bm{\eta}, \bm{\zeta} \in \bm{\mathcal{M}}_{h}^k$, we write
\begin{subequations}
\label{innerProducts}
\begin{alignat}{3}
& (\bm{w},\bm{v})_{\mathcal{T}_h} && = \sum_{K \in \mathcal{T}_h} (\bm{w}, \bm{v})_K && = \sum_{K \in \mathcal{T}_h} \int_{K} \bm{w} \cdot \bm{v} , \\
& (\bm{W},\bm{V})_{\mathcal{T}_h} && = \sum_{K \in \mathcal{T}_h} (\bm{W}, \bm{V})_K && = \sum_{K \in \mathcal{T}_h} \int_{K} \bm{W} : \bm{V} , \\ 
& \left\langle \bm{\eta}, \bm{\zeta} \right\rangle_{\partial \mathcal{T}_h} && = \sum_{K \in \mathcal{T}_h} \left\langle \bm{\eta},\bm{\zeta} \right\rangle_{\partial K} && = \sum_{K \in \mathcal{T}_h} \int_{\partial K} \bm{\eta} \cdot \bm{\zeta} , 
\end{alignat}
\end{subequations}
where $:$ denotes the Frobenius inner product of two matrices.

\subsection*{\label{s:numDisApp}Hybridized DG discretization}

The hybridized DG discretization of the (unfiltered) unsteady compressible Navier-Stokes equations reads as follows: Find $\big( \bm{q}_h(t),\bm{u}_h(t), \widehat{\bm{u}}_h(t) \big) \in \bm{\mathcal{Q}}_h^k \times \bm{\mathcal{V}}_h^k \times \bm{\mathcal{M}}_h^k$ such that
\begin{subequations}
\label{e:hDG_NS}
\begin{alignat}{2}
\label{e:hDG_NS0}
\big( \bm{q}_h, \bm{r} \big) _{\mathcal{T}_h} + \big( \bm{u}_h, \nabla \cdot \bm{r} \big)  _{\mathcal{T}_h} -  \big< \widehat{\bm{u}}_h, \bm{r} \cdot \bm{n} \big> _{\partial \mathcal{T}_h}  & =  0 , \\
\label{e:hDG_NS1}
\Big( \frac{\partial \, \bm{u}_h}{\partial t}, \bm{w} \Big)_{\mathcal{T}_h} - \Big( \bm{F}(\bm{u}_h) + \bm{G}(\bm{u}_h,\bm{q}_h) , \nabla \bm{w} \Big) _{\mathcal{T}_h}  +  \left\langle \widehat{\bm{f}}_h(\widehat{\bm{u}}_h,\bm{u}_h) + \widehat{\bm{g}}_h(\widehat{\bm{u}}_h,\bm{u}_h,\bm{q}_h), \bm{w} \right\rangle_{\partial \mathcal{T}_h}  & = 0 , \\
\label{e:hDG_NS2}
\left\langle \widehat{\bm{f}}_h(\widehat{\bm{u}}_h,\bm{u}_h) + \widehat{\bm{g}}_h(\widehat{\bm{u}}_h,\bm{u}_h,\bm{q}_h), \bm{\mu} \right\rangle_{\partial \mathcal{T}_h \backslash \partial \Omega} + \left\langle \widehat{\bm{b}}_h(\widehat{\bm{u}}_h,\bm{u}_h,\bm{q}_h), \bm{\mu} \right\rangle_{\partial \Omega} & =  0 , \\
\intertext{for all $(\bm{r},\bm{w}, {\bm{\mu}}) \in \bm{\mathcal{Q}}^k_h \times \bm{\mathcal{V}}^k_h \times \bm{\mathcal{M}}_{h}^k$ and all $t \in [0,t_f)$, as well as}
\label{e:hDG_NS3}
\big( \bm{u}_{h}|_{t=0} - \bm{u}_0 , \bm{w} \big) _{\mathcal{T}_h} & =  0 , 
\end{alignat}
\end{subequations}
for all $\bm{w} \in \bm{\mathcal{V}}_h^k$. Here, $\bm{n}$ denotes the unit normal vector pointing outwards from the elements, $\widehat{\bm{b}}_h$ is the boundary condition term (whose precise definition depends on the type of boundary condition), and $\widehat{\bm{f}}_h$ and $\widehat{\bm{g}}_h$ are the inviscid and viscous numerical fluxes defined as
\begin{subequations}
\label{numericalFlux}
\begin{alignat}{2}
\label{e:numericalInviscidFlux}
& \widehat{\bm{f}}_h(\widehat{\bm{u}}_h , \bm{u}_h ) = \bm{F}(\widehat{\bm{u}}_h) \cdot \bm{n} + \bm{\sigma}(\widehat{\bm{u}}_h , \bm{u}_h ; \bm{n}) \cdot ( \bm{u}_h - \widehat{\bm{u}}_h ) , \\
\label{e:numericalViscousFluxNS}
& \widehat{\bm{g}}_h(\widehat{\bm{u}}_h , \bm{u}_h , \bm{q}_h ) = \bm{G}(\widehat{\bm{u}}_h , \bm{q}_h) \cdot \bm{n} . 
\end{alignat}
\end{subequations}
We note that this form of the numerical flux does not involve an explicit Riemann solver on the element faces. 
Instead, it is the so-called stabilization matrix $\bm{{{\sigma}}} \in \mathbb{R}^{m \times m}$ that implicitly defines the Riemann solver in hybridized DG methods. The interested reader is referred to \cite{Fernandez:AIAA:17a} for a discussion on the relationship between the stabilization matrix and the resulting Riemann solver. The hybridized DG discretization of the unsteady compressible Euler equations is obtained by dropping Eq. \eqref{e:hDG_NS0} and the viscous terms in Equations \eqref{e:hDG_NS1}$-$\eqref{e:hDG_NS2}. Finally, in the context of explicit LES, we augment the viscous stress tensor and heat flux in \eqref{e:closuresNS} with the modeling terms
\begin{equation}
\label{e:closuresNS_SGS}
\tau_{ij}^{SGS} = \rho \, \nu_e \, \bigg( \frac{\partial v_i}{\partial x_j} +\frac{\partial v_j}{\partial x_i} - \frac{2}{3}\frac{\partial v_k}{\partial x_k}\delta_{ij} \bigg) + \tau_{iso}^{SGS} \, \delta_{ij} , \qquad \qquad f_j^{SGS} = - \, \frac{\rho  \, \nu_e \, c_p}{Pr_e} \, \frac{\partial T}{\partial x_j} , 
\end{equation}
where $\nu_e$ is the kinematic eddy viscosity, $\tau_{iso}^{SGS}$ the isotropic part of the subgrid-scale stress tensor, and $Pr_e$ the SGS eddy Prandtl number, and are computed using the subgrid-scale models discussed in Section \ref{s:summaryCases}.  
Additional details on the hybridized DG methods for compressible flows are presented in \cite{Fernandez:17a}. 



\end{document}